%% file: main.tex
\documentclass[letterpaper,twocolumn,10pt]{article}
\usepackage{hyperref}
\usepackage{breakurl}
\usepackage{usenix-2020-09}
\usepackage{tikz}
\usepackage{xspace}
\usepackage{amsmath}
\usepackage{cleveref}
\crefname{section}{}{\S\S}
\crefdefaultlabelformat{\S#2#1#3}
\usepackage[normalem]{ulem}
\usepackage{enumitem}
\setlist{nolistsep}
\usepackage{comment}
\usepackage{booktabs}
\usepackage{multirow}
\usepackage{caption}
\usepackage{subcaption}
\usepackage{balance}
\usepackage{algorithm}
\usepackage{algpseudocode}

\usepackage[small, compact]{titlesec}
\titlespacing*{\section}{0pt}{8pt}{4pt}
\titlespacing*{\subsection}{0pt}{4pt}{2pt}
\titlespacing*{\subsubsection}{0pt}{2pt}{1.5pt}

\newcommand{\aditya}[1]{{\color{red} aditya: #1}}

\newcommand{\sys}{\texttt{BlockLLM}\xspace}
\newcommand{\block}{block\xspace}
\newcommand{\blocks}{blocks\xspace}
\newcommand{\Blocks}{Blocks\xspace}
\newcommand{\Block}{Block\xspace}
\newcommand{\PEFT}{PE\xspace}
\newcommand{\FPFT}{FF\xspace}
\newcommand*\circled[1]{\tikz[baseline=(char.base)]{
            \node[shape=circle,draw,inner sep=0.5pt] (char) {#1};}}
\begin{document}

\date{}
% \title{\sys: Multi-tenant Finer-grained Serving for Large Language Models}
\title{\sys: Multi-tenant Finer-grained Serving for Large Language Models}
% \author{Paper \#1770, 12 pages plus references}
\author{
{\rm Bodun Hu}\\
The University of Texas at Austin
\and
{\rm Jiamin Li}\\
Microsoft Research
\and
{\rm Le Xu\thanks{Work done at the University of Texas at Austin.}}\\
The University of Texas at Austin
\and
{\rm Myungjin Lee}\\
Cisco Research
\and
{\rm Akshay Jajoo}\\
Cisco Research
\and
{\rm Geon-Woo Kim}\\
The University of Texas at Austin
\and
{\rm Hong Xu}\\
The Chinese University of Hong Kong
\and
{\rm Aditya Akella}\\
The University of Texas at Austin
% copy the following lines to add more authors
% \and
% {\rm Name}\\
%Name Institution
} % end author

\maketitle
% \twocolumn[\begin{@twocolumnfalse}

% \begin{centering}
% \vspace{0.1cm}
% {\Large \bf \sys: Finer-Grained Large Language Model Serving in Multi-tenant Clouds
% \\}
% \vspace{0.15cm}
% Paper \#1770, 12 pages plus references

% \end{centering}
% \vspace{\baselineskip}
% \vspace{0.6cm}

% \end{@twocolumnfalse}]

\input{abstract}
\input{introduction}
\input{motivation}
\input{design}
\input{modelzoo}

\input{onlineserving}
\input{implementation}

\input{evaluation}
\input{discussion}
\input{related}
\input{conclusion}
\bibliographystyle{plain}
\bibliography{main}
\balance
\end{document}

%% file: abstract.tex
\begin{abstract}
The increasing demand for Large Language Models (LLMs) across various applications
has led to a significant shift in the design of deep learning serving systems.
Deploying LLMs, particularly in multi-tenant environments, poses substantial
challenges due to their high computational and memory demands. We introduce \sys,
a serving system that leverages component sharing among
fine-tuned LLM models to provide an efficient and flexible solution for LLM
workloads. \sys partitions models into finer-grained \blocks, enabling the
reuse of model components and independent provisioning to improve
computation efficiency. \sys comprises an offline \block zoo for storing
\blocks and an online system to serve requests through chains of \blocks.
It offers multi-fold flexibilities: (1) Adaptive assembly of \blocks 
on-the-fly through equivalence evaluation among blocks in the zoo; (2) Per-\block batch size configuration and best-effort KV cache coordination at the individual \block level; (3) Speculative execution and locality-aware \block placement to reduce communication costs from dynamic \block resource allocation. Our evaluation shows that \sys reduces memory and storage footprints and improves computational efficiency, outperforming existing serving approach in 95\%ile latency and GPU utilization by 33.5\% and 20.1\%, respectively, with minimal impact on accuracy.
\end{abstract}

%% file: introduction.tex
\section{Introduction}
\label{sec:intro}
The rise of Large Language Models (LLMs) marks a transformative milestone in deep learning. Their unprecedented abilities in natural language processing~\cite{min2023recent}, from language translation~\cite{conneau2019cross} to question-answering~\cite{ChatGPT} and complex reasoning~\cite{lewkowycz2022solving, wei2022chain}, are reshaping technological frontiers. Yet, the strength of LLMs---rooted in their vast parameter spaces---comes at the price of significant computation costs~\cite{chowdhery2022palm}. The deployment of LLMs is an arduous endeavor, often necessitating the use of hundreds or even thousands of extremely expensive computation devices like GPUs, thus imposing great challenges on unleashing LLM's full potential~\cite{shoeybi2019megatron,aminabadi2022deepspeed}. 

% With the growing reliance on LLMs, 
Fine-tuning recently has emerged as a critical technique to efficiently adapt foundation models to handle various downstream tasks across different domains~\cite{houlsby2019parameter,ding2023parameter}. Fine-tuning updates existing parameters or introduces additional parameters using domain-specific data. As LLMs become ubiquitous, the challenge of serving these fine-tuned models from a multi-tenant cluster---whether in a private cloud for first-party services or a public cloud for third-party users---has become increasingly significant. %become increasingly complex. %LLM serving systems must not only fulfill the performance needs of fine-tuned model inference but also orchestrate the intensive resource needs of diverse applications. % to maximize utilization.

% The wide adoption of LLM necessitates an efficient multi-tenant serving system tailored for such workloads. Beyond fulfilling latency requirements, our goal is to elevate throughput and resource utilization in a cluster where diverse LLM applications are present.
\begin{comment}
Recently extensive studies on serving LLM workloads tackle issues such as the acceleration of matrix computations~\cite{dao2023flashattention, FasterTransformer} and optimizing tensors' memory allocation~\cite{kwon2023efficient} to enabling parallel decoding to circumvent the limitations imposed by auto-regressive models that process a single token at a time~\cite{fu2023lookahead, miao2023specinfer}. Despite these advancements, efficiently managing a diverse range of LLMs in multi-tenant clusters has yet to be fully addressed. \aditya{is the problem we are addressing about "management"? What does that mean?} The prohibitive costs of hardware prevent the cluster from expanding liberally. \aditya{without explaining what we are targeting, unclear what expanding means and why it is not good} Therefore, LLM serving system should extend beyond internal model optimization \aditya{elaborate? are you referring to the current techniques listed in the first sentence?}, encompassing efficient multi-model \aditya{broken sentence?} and resource management---a challenge that stands out as particularly pressing. \aditya{this paragraph is confusing and weak. Also English seems broken. It's an important paragraph! We need to clearly say what is the problem with existing techniques.}
\end{comment}
Recent serving systems and approaches focus on model-internal optimization, such as accelerating matrix computations~\cite{dao2023flashattention, FasterTransformer}, minimizing device memory fragmentation~\cite{kwon2023efficient}, and enabling parallel decoding in the auto-regressive token generation process~\cite{fu2023lookahead, miao2023specinfer}. While these advancements significantly accelerate inference of individual LLMs, the challenge remains to serve multiple tenants' fine-tuned models in a shared cluster. Each LLM requires substantial resources, and the hardware cost makes it prohibitive to liberally expand the cluster to add dedicated resources for each tenant. Thus, the question is: {\em how can LLM serving meet the latency goals of tenants' fine-tuned models while also ensuring high throughput and optimal utilization cluster-wide?} 

% \ljm{cluster cannot expand liberally is mentioned here => indicating resource-constraint environment implicitly}

We introduce \sys, an efficient and flexible serving system tailored for fine-tuned LLMs in a multi-tenant cluster. \sys enables LLM model component reuse by breaking down an LLM into smaller and shareable components which we call ``\blocks''. The central observation forming the basis for \sys is that fine-tuning offers the opportunity for {\em sharing}. Parameter-efficient fine-tuned models have an inherent modular architecture where parameters from specific layers are added or altered. Our observations with full-parameter fine-tuned models indicate that certain model components produce outputs with high similarity, suggesting their potential for sharing as well. \Block sharing reduces both memory and storage demands. Furthermore, by reallocating memory---previously consumed by redundant parameters---towards supporting larger data batch size, \sys achieves higher computation efficiency and overall throughput. 

\sys consists of an offline ``\block zoo''---a repository of LLM \blocks, and an online serving system capable of dynamically scheduling \blocks to serve various applications (Figure~\ref{fig:sys_arch}). \sys brings three immediate opportunities. (1) It can support adaptive serving, which enables on-the-fly assembly of \blocks for each request, moving beyond the traditional static chain of \blocks. (2) Each \block can be individually configured, allowing for customized batch sizes and autonomous handling of queued requests. (3) \sys enables dynamic resource allocation, with each \block capable of independently scaling and being placed without the constraints of model-level boundaries.

Several challenges arise in realizing \sys with full flexibility. First, the feasibility of adaptive serving hinges on an in-depth understanding of the equivalence among \blocks. This necessitates a mechanism to establish inter-\block relationship within the \block zoo. Moreover, LLMs store intermediate results in KV cache~\cite{pope2023efficiently} to avoid recalculation due to their auto-regressive nature. The independent configuration of each \block leads to overlapping lifecycles of different requests, demanding stateful coordination to maintain the coherence of the KV cache across concurrent requests. %\aditya{first mention of KV cache in the intro. Need to fix} 
Lastly, resource allocation without model-level boundaries brings extra inter-\block communication, thereby imposing additional latency costs that are not easily mitigated. In \sys, we propose integrated solutions for these challenges. 

\textit{First}, to ascertain the equivalence of \blocks, we compare the output distribution of vocabulary probabilities, which serves as an indicator for functional equivalence between transformer layers. Furthermore, we investigate methods to facilitate the reuse and sharing of LLM components while minimizing any potential impact on accuracy.
%\ml{Why not applying comparing the output distribution of vocabulary probabilities for models of both the same architecture and different ones? If we use parameter similarity for the same architecture, I think we should show that using the parameter similarity doesn't affect the accuracy or the output similarity.} \bd{Updated this paragraph}
Inspired by the techniques of merging computer vision models~\cite{pan2023stitchable}, we design a generalizable stitching \block as an intermediary to route requests among these \blocks so that the chain of \blocks to serve each request can be adaptively adjusted online.

\textit{Second}, to manage the KV cache generated by different requests, \sys employs a principled approach to request dispatching. The per-block configuration introduces complexity in handling the KV cache, particularly when requests are assigned to devices lacking the corresponding cache. This scenario necessitates either cache communication or recalculation. We analyze the trade-off between recalculation and I/O costs to determine priorities for different candidate \block instances, including those specified in the chain and its equivalent \blocks. We adopt a best-effort KV coordination policy, prioritizing devices that already possess the KV cache when dispatching requests.

\textit{Third}, to address the inherent latency costs associated with inter-\block communication, \sys integrates speculative execution for identified bottleneck \blocks, enabling the processing of potential request pathways in advance. We use surrogate models to predict the outputs of \blocks, facilitating computation ahead of actual output. We also implement a locality-aware strategy for \block placement, positioning interdependent \blocks in close proximity, preferably within the same server. This approach minimizes the frequency of costly inter-server communications and leverages the high-speed intra-server connections to expedite data transfer.

% \aditya{drop this paragraph - it is redundant}
% Putting everything together, we design \sys, a new LLM serving system that realizes \block-granularity provision to improve cluster efficiency. \sys has a \block zoo, equipped with an automated tool for on-demand model partitioning and \block reusing. It also evaluates \block equivalence to support adaptive serving. \sys's online scheduler orchestrates the resource allocation and placement of the \blocks to reduce communications and applies speculative execution on select \blocks, expediting the inference pipeline. \sys's agents, deployed on each device, oversee the processing of request batches through a principled scheduling policy to dispatch requests and their associated KV cache if necessary. 

We summarize our contributions as follows:

\begin{itemize}[noitemsep,topsep=0pt,leftmargin=*]
    \item We leverage the LLM fine-tuning characteristic and demonstrate the benefits of finer-grained LLM serving. This approach reduces memory and storage demands while optimizing computation resource utilization.
    \item We construct a \block zoo that stores LLMs in finer granularity to enable \block reuse and establishes the equivalence among these \blocks to facilitate adaptive online serving.
    \item We build an online serving system to improve cluster throughput. It strategically coordinates the KV cache by best-effort request dispatching and minimizes communication overhead with bottleneck-pinpointing speculative execution and locality-aware placement.
    \item We implement and evaluate \sys in a 12-A100 cluster. \sys achieves comparable median latency with traditional per-model provisioning and reduces the 95\%ile latency by 33.5\%. The overall average throughput is elevated by 1.71x. GPU SM efficiency is improved by 20.1\%. %Evaluation using testbed experiments validates its effectiveness.
\end{itemize}

%% file: motivation.tex
\section{Background and Motivation}
\label{sec:motivation}
\subsection{Background}
\label{subsec:background}

\begin{figure}[t]
    \centering
    \includegraphics[width=\columnwidth]{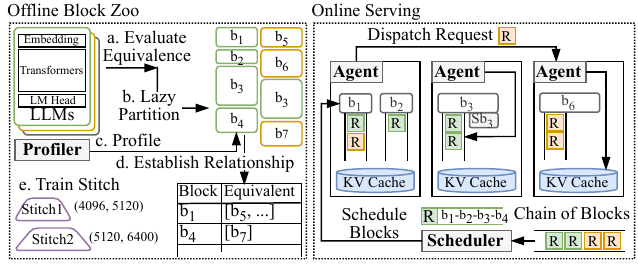}
    \vspace{-7mm}
    \caption{System architecture. $b_x$ are \blocks. `R' are requests. $SB_3$ is the surrogates of \Block $b_3$.}
    \vspace{-7mm}
    \label{fig:sys_arch}
\end{figure}

\noindent\textbf{LLM serving.} LLM-based applications have grown to become major workloads in GPU clusters, creating significant serving challenges. Large storage is required to store the model checkpoints and engines, and numerous powerful GPUs with ample memory are required to load and execute the models.~\cite{sheng2023flexgen,pope2023efficiently}. The auto-regressive nature of LLMs necessitates caching KV matrices to avoid redundant recalculations~\cite{radford2019language}. Each request's KV cache grows as the generation process continues. Moreover, when models are inevitably deployed in a distributed manner, a high-capacity network is indispensable to transfer the intermediate activations. Such a burden is particularly pronounced in multi-tenant clusters, which cater to diverse LLM applications, each with its dedicated models and performance objectives. 

\noindent\textbf{Foundation models and fine-tuning.} LLM training follows transfer learning as a guiding principle. A foundation model is initially pre-trained on a broad spectrum of general data, which is then customized for specific applications through fine-tuning. This process involves introducing additional parameters or modifying existing ones within the model using task-specific datasets. Fine-tuning entails updating either the entire model, known as full-parameter fine-tuning (\FPFT)~\cite{vicuna2023}, or altering a small subset of additional or modified parameters, referred to as parameter-efficient fine-tuning (\PEFT)~\cite{peft,he2021towards}. \PEFT examples include LoRA~\cite{hu2021lora}, BitFit~\cite{zaken2021bitfit}, Adapter~\cite{he2021effectiveness}, and Prefix-Tuning~\cite{li2021prefix}.
\begin{comment}
The prevailing approach is for large organizations with ample data and computation resources to develop and periodically update a few foundation models (e.g., LLaMA, GPT). Other users then fine-tune these models with their data to create tailored solutions for various downstream tasks. The update of foundation models is less frequent than the fine-tuned models due to the dataset availability and considerable costs of a full set training cycle.
\end{comment}

\noindent\textbf{Problem statement.} The wide adoption of LLM necessitates an efficient multi-tenant serving system tailored for such workloads. Beyond fulfilling latency requirements, our goal is to elevate throughput and resource utilization in a cluster where diverse LLM applications are present. 
% \aditya{a version of this paragraph can go into the third paragraph of the intro}

\subsection{Key Idea: Finer-grained Serving}
\label{subsec:block_serving}
To enhance the efficiency of LLM serving systems, we advocate a finer-grained approach. Rather than treating each model as an indivisible unit, we propose \sys,  which partitions LLMs into smaller components provisioned independently. These components are envisioned as ``\blocks''. We could maintain only one copy of a \block even if it is used by multiple models. The immediate benefit is the decreased demand for memory and storage. 
\begin{figure}[t]
    \centering
    \includegraphics[width=0.9\columnwidth]{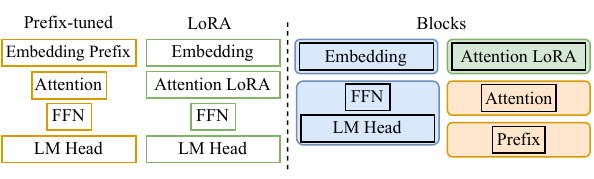}
    \vspace{-4mm}
    \caption{Example of \blocks for two models fine-tuned from the same foundation. We show one Transformer layer for simplicity.}
    \vspace{-3mm}
    \label{fig:block_def}
\end{figure}

\begin{figure}[t] 
    \begin{subfigure}{0.50\columnwidth}
        \includegraphics[width=\columnwidth]{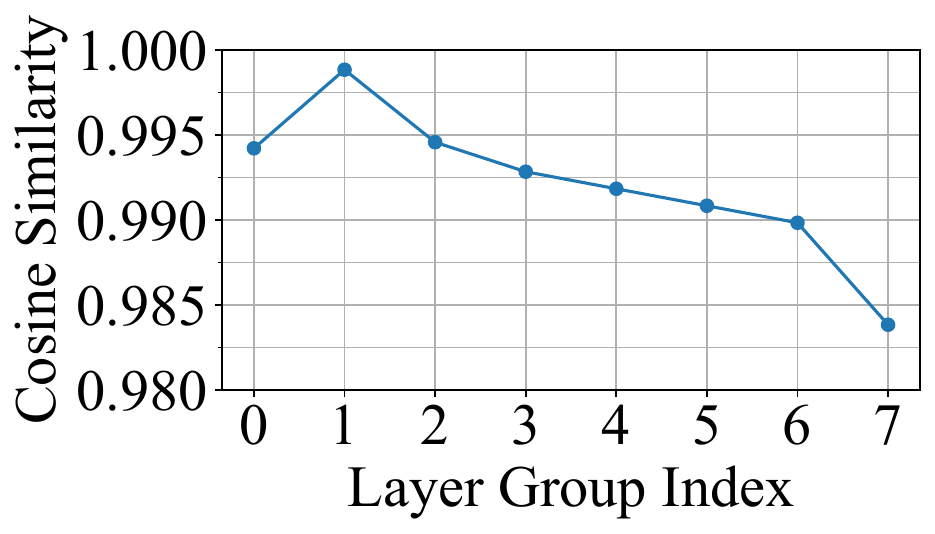}
        \centering
        \vspace{-5mm}
        \caption{LLaMA 7B and LLaMA 13B.}
        \label{fig:llama_difff_7_13_similarity}
    \end{subfigure}%
    \hfill
    \begin{subfigure}{0.47\columnwidth}
        \includegraphics[width=\columnwidth]{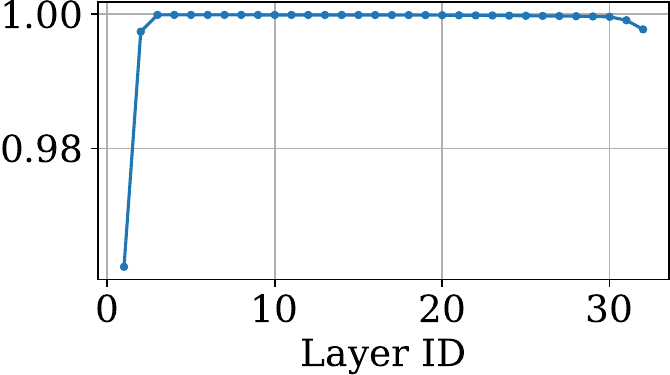}
        \vspace{-5mm}
        \centering
        \caption{LLaMA 7B and Vicuna 7B.}
        \label{fig:llama_vicuna_output_sim}
    \end{subfigure}
    \caption{Output %cosine
    similarity between Transformer layer groups of two models.}
    \label{fig:output_cos_similarity}
    \vspace{-7mm}
\end{figure}

\PEFT models are ideal for this approach due to their inherent modular architecture. As shown in Figure~\ref{fig:block_def}, two models could be divided into five \blocks, with blue \blocks being shared components. Moreover, our analysis of \FPFT models reveals that different \blocks share highly similar outputs. 
%\ml{Discrepancy. The previous sentence talks about output similarity. The next sentence talks about parameter similarity.} \bd{addressed} 

Figure~\ref{fig:output_cos_similarity} presents the output vocabulary distribution similarity of each Transformer layer between LLaMA 7B and LLaMA13B, as well LLaMA 7B and Vicuna 7B(\FPFT from LLaMA), with an average similarity of 0.9927 and 0.998, respectively. This observation underpins the feasibility of reusing \blocks across different \FPFT models, thereby enhancing resource efficiency by allowing device memory to hold more critical data.

\noindent\textbf{Opportunities.} \sys consists of an offline \block zoo (\Cref{sec:blockzoo}) and an online serving system (\Cref{sec:serving}). We enable \block reuse in \block zoo and establish connections among \blocks. In the serving phase, \sys deploys the \blocks onto the cluster and assembles a chain of \blocks to serve each request. Each \block instance is provisioned independently. \Block-granularity provision brings three significant opportunities for enhancing serving flexibility. 

% \ljm{aditya: should be opportunity instead of benefits, and mention there is a challenge for every opportunity}

\noindent\textbf{O1: Adaptive serving.} Deployment and serving can be highly adaptive. \Blocks allow for dynamic assembly into a variety of model configurations. Serving a request is not constrained to a predetermined chain of \blocks. Instead, it can be dynamically adjusted in real-time, based on the deployment state, to improve efficiency. The key to realizing this fully adaptive serving hinges on accurately determining whether the \blocks are functionally equivalent (\Cref{sec:blockzoo}, \Cref{subsec:schedule}). % \aditya{give a forward pointer}

\noindent\textbf{O2: Per-\block configuration.} Each \block can have its unique configuration. First, the batch size of each \block could be individually configured. Intuitively, a \block shared by multiple applications should have a larger batch size to efficiently handle the collective load. Second, each \block maintains its own request queue and manages the request state independently, thus autonomously deciding how to process the requests. A critical consideration in this configuration is how auto-regressive models can be integrated into this paradigm, where the KV cache should be managed in a cohesive approach (\Cref{subsec:statful}). 

\noindent\textbf{O3: Dynamic resource allocation.} \sys enhances resource allocation
%with its ability to independently scale
by scaling each \block independently. \Blocks that are frequently accessed or computationally intensive can dynamically be allocated with more resources. It also liberates \block placement from the constraints associated with monolithic model architectures. However, it does introduce a challenge in the form of increased communication costs. Effectively mitigating these costs is crucial to achieving truly versatile resource allocation within \sys (\Cref{subsec:speculative}, \Cref{subsec:schedule}). % \aditya{forward pointer}

% \aditya{if you have forward pointers above then you can drop the following IMO}
% We will thoroughly analyze the challenges that emerged from these opportunities in~\Cref{subsec:challenges} and discuss other opportunities to be explored in~\Cref{sec:discussion}.

\subsection{Comparison with Existing Work}
\label{subsec:existing_work}
%\noindent\textbf{Existing serving systems.} 
%Existing serving systems for LLM
%workload follow the practice of serving traditional DNNs. 
Existing serving systems for serving LLM workloads utilize the same techniques as traditional DNN serving. Each model operates
within a dedicated engine and is provisioned as a monolithic unit. An
alternative approach is multi-tasking. A typical example is
S-LoRA~\cite{sheng2023s}, which exploits memory paging to pack multiple LoRAs
within one engine. Another line of optimizations focuses on the model
internals---like accelerating Attention layer computations
(FlashAttention~\cite{dao2023flashattention}), managing KV cache
(PagedAttention~\cite{kwon2023efficient} in vLLM), and efficiently scheduling
varying sequence lengths of requests (Orca~\cite{yu2022orca}). These efforts are
complementary to \sys. \sys tackles the inefficiency of serving LLM workloads in
a multi-tenant cluster by leveraging the characteristics of LLM fine-tuning.
\Block-granularity provisioning improves the serving efficiency from two
perspectives.
\begin{figure}[t]
    \begin{minipage}{0.66\columnwidth}
        \resizebox{\columnwidth}{!}{\begin{tabular}{@{}llc@{}}
        \toprule
        Model & \PEFT & \% of Shared Params. \\ \midrule
        \multirow{3}{*}{LLaMA 7B} & LoRA (Green) & 99.94    \\
                         & Adapter (Purple) &   92.62        \\
                         & Prefix (Orange) &   99.88  \\
        \multirow{3}{*}{GPT-NeoX-20B}  &  LoRA    &   99.52                    \\
                         &  Adapter    &  92.52                     \\
                         &  Prompt    &   99.98                    \\ \bottomrule
        \end{tabular}}
        \captionof{table}{Percentage of shared parameters of different \PEFT techniques.}
        \label{table:peft_percent}
    \end{minipage}
    \hfill
    \begin{minipage}{0.3\columnwidth}
        \centering
        \includegraphics[width=\columnwidth]{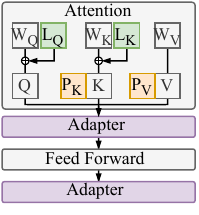}
        \vspace{-7mm}
        \caption{Architectural changes of \PEFT.}
        \label{fig:peft_arch}
    \end{minipage}
    \vspace{-5mm}
\end{figure}
\begin{figure}
    \centering
    \includegraphics[width=0.7\columnwidth]{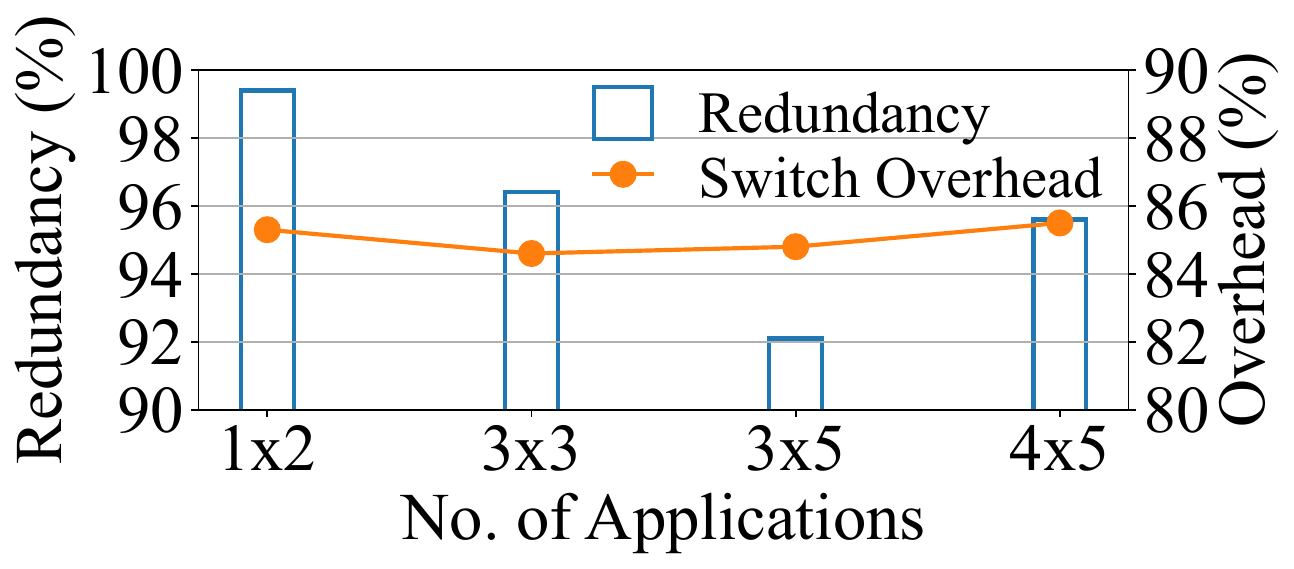}
    \vspace{-4mm}
    \caption{Redundancy and switching overhead in the existing serving system. x$\times$y indicates we have x foundation models and y models fine-tuned from each foundation. Therefore, xy models in total.}
    \vspace{-5mm}
    \label{fig:redundancy_overhead}
\end{figure}

\noindent\textbf{Reduced memory and storage.} Models fine-tuned from the same foundation model inherit a significant number of shared parameters. We detail various fine-tuning techniques and the extent of parameter modification in Table~\ref{table:peft_percent} and illustrate the architectural changes in Figure~\ref{fig:peft_arch} with shared parameters depicted as gray boxes. Existing serving systems ignore this aspect, leading unfortunately to the redundant storage of identical model pieces. In Figure~\ref{fig:redundancy_overhead}, we present the percentage of the model parameters being redundant. With more diverse applications served in a cluster, the size of redundancy is more significant. When the cluster hosts 15 LLM-based applications that are adapted from three foundation models (3rd bar), the 92.1\% redundancy takes up to $\sim$147GB. In terms of memory-related operations, this redundancy becomes particularly instrumental, where models are frequently swapped in and out in response to requests from different applications. A considerable portion of the switching overhead (85.4\% with 20 applications) is wastefully spent on unnecessary replacement of the same parameters. With \sys, the switching overhead is reduced to at most 12.2\% with 9 applications and 16.5\% with 20 applications.

\begin{table}[t]
\resizebox{\columnwidth}{!}{\begin{tabular}{@{}cllllll@{}}
\toprule
\multirow{2}{*}{\begin{tabular}[c]{@{}l@{}}No. of \\ Apps\end{tabular}} & \multicolumn{3}{c}{Per-model Provision}          & \multicolumn{3}{c}{\sys}                \\ \cmidrule(l){2-7} 
                            & Latency & Throughput & Utilization & Latency & Throughput & Utilization \\ \midrule
3    &   5.4    &  6.6       &   64.4\%        &  4.6  &   8.9    &     76.5\%      \\
6    &   7.6    &  8.2       &   75.1\%        &  5.2  &  12.3    &     79.4\%             \\
9    &  14.4    &  7.2       &   76.6\%        &  9.9  &  14.5    &     85.3\%        \\
12   &  17.6    &  6.9       &   74.4\%        & 12.4  &  15.6    &     89.6\%       \\ \bottomrule
\end{tabular}}
\vspace{-3mm}
\caption{Comparison of average latency (mins/request), throughput (tokens/second), and GPU utilization (SM efficiency) when the number of applications increases. GPU utilization is measured until the last request is completed. }
\vspace{-3mm}
\label{table:baseline_ours_motivation}
\end{table}

\noindent\textbf{Improved computation efficiency.} Serving each model within a standalone engine presupposes a complete execution cycle. As such, per-model provisioning prevents the sharing of computations among models, even when they contain identical components. It is inefficient in a multi-tenant cluster where many different applications are served. Table~\ref{table:baseline_ours_motivation} shows the average latency, throughput, and GPU utilization of a small cluster of four GPUs where 40 requests from three applications are served. When using per-model provisioning, the throughput and GPU utilization are low because the batch size is small, and the devices hosting less popular applications are severely under-utilized. Alternatively, with \sys, we could achieve a 34.8\% higher throughput. The gain primarily stems from three aspects: (1) the reduced overhead by loading smaller model pieces, (2) higher computation efficiency due to larger batch sizes for shared \blocks (\textbf{O2}), (3) effective scaling of popular \blocks onto the idle computation resources originally reserved by less popular applications (\textbf{O3}). We also measure the performance improvement when the number of applications the cluster hosts grows from 3 to 12 and the number of requests grows proportionally. The throughput improvement of 12 applications is 1.91x of the throughput improvement when hosting three applications. 
% \aditya{1.91x of the improvement, or of the throughput?} 

%% file: design.tex
\section{\sys Design}
\label{sec:design}
%We provide an overview of \sys and explain its three design challenges.
In this section, we provide an overview of the \sys's architecture and workflow, along with a discussion of its three key design challenges.
\subsection{System Overview}
\label{subsec:overview}
\noindent\textbf{Architecture.} Figure~\ref{fig:sys_arch} depicts the system architecture of \sys. In the preparatory offline phase, \sys hosts a repository—termed ``\block zoo''---that houses the LLMs in \blocks. It is equipped with an automated mechanism that partitions the models into \blocks and ascertains the equivalence among existing \blocks. Additionally, the \block zoo integrates a profiler to record the performance metrics and trade-offs pertinent to each \block. In the real-time online phase, \sys has a scheduler that orchestrates resource allocation and placement of \blocks and processes the requests. It strategically schedules \blocks onto devices, denoted as ``\block instance''. A \sys agent resides on every device in the cluster. It maintains and monitors the \block instances and the associated request queues. It is responsible for handling the requests, including managing the KV cache and transferring outputs to other \block instances.

\noindent\textbf{Workflow.} 
%During serving, the workflow is as follows.
When a new request arrives, the scheduler assigns a chain of \blocks based on the application it belongs to. The scheduler initially determines whether there is an available instance of the first \block in the chain. If available, the request is forwarded to the target \block instance. Otherwise, the scheduler either identifies a suitable alternative among the existing equivalent \block instances or deploys a new \block instance on an available device. Upon receipt of the request, the device's agent directs it to the designated \block instance for processing. Then, once the execution is completed, the agent is responsible for directing the request through the subsequent \block in the chain. The process ends with the completion of the request signified by an EOS token, whereupon the final agent relays the output back to the scheduler, thus concluding the response cycle of the request.

\subsection{Design Challenges}
\label{subsec:challenges}

\noindent\textbf{C1: Evaluate different-sized \block equivalence.}  
To achieve adaptive serving, establishing the equivalence between \blocks is necessary so that chains of \blocks for each request can be adjusted on the fly. \Block-granularity model zoo offers improved manageability compared to full models due to their significantly smaller parameter size. Nonetheless, establishing such equivalence remains a complex issue. Despite the widespread adoption of the Transformer architecture in most existing LLMs, they often vary in size-related parameters like the embedding size. While metrics such as cosine similarity or L2 norm can be employed to assess the equivalence between \blocks of identical architecture, evaluating \blocks of differing architectures for equivalence is far from straightforward. Furthermore, even with an understanding of their similarities, it is not feasible to directly route a request to a \block that differs in embedding size.
\begin{figure}[t]
    \centering
    \begin{subfigure}{.46\columnwidth}
        \includegraphics[width=\columnwidth]{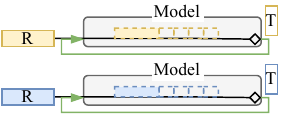}
        \vspace{-6mm}
        \caption{Per-model provision.}
        \label{fig:kv_cache}
    \end{subfigure}%
    \hfill
    \begin{subfigure}{.46\columnwidth}
        \includegraphics[width=\columnwidth]{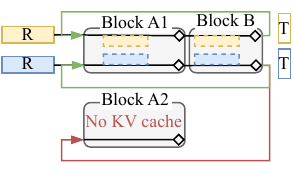}
        \vspace{-6mm}
        \caption{\Block provision.}
        \label{fig:stateful_kv_cache}
    \end{subfigure}
    \vspace{-3mm}
    \caption{Illustration of coordinating KV cache when using \block-granularity provisioning }
    \label{fig:kv_challenge}
    \vspace{-6mm}
\end{figure}

\noindent\textbf{C2: Coordinate stateful KV cache.}
% \ljm{\hx{\noindent\textbf{C2: Multi-batch KV cache.}} I don't think it is multi-batch KV cache. We are not optimizing the memory allocation of multiple KV cache. We need to maintain and redirect (if necessary) KV cache of each request batch.} 
In \sys, sophisticated stateful serving is indispensable for auto-regressive LLMs. Existing serving systems operate by dedicating each model instance to process a single batch of requests until completion. This approach allows GPU devices to retain a singular, specific KV cache for the batch they are processing. However, \sys introduces a more complex scenario where each \block is configured independently. Each  \block instance may concurrently process multiple request batches over time, each generating its distinct KV cache. This interleaved processing approach, coupled with the auto-regressive nature of the tasks, implies that a request batch requiring subsequent iterations needs access to its original KV cache. However, there's no guarantee that the same device that holds the KV cache will be available to process additional iterations of that batch. Figure~\ref{fig:kv_challenge} illustrates this challenge, showcasing how the blue request could encounter the problem of no available KV cache if dispatched to \Block instance A2 under \sys's provision choice.
\begin{figure}[t]
    \centering
    \includegraphics[width=0.6\columnwidth]{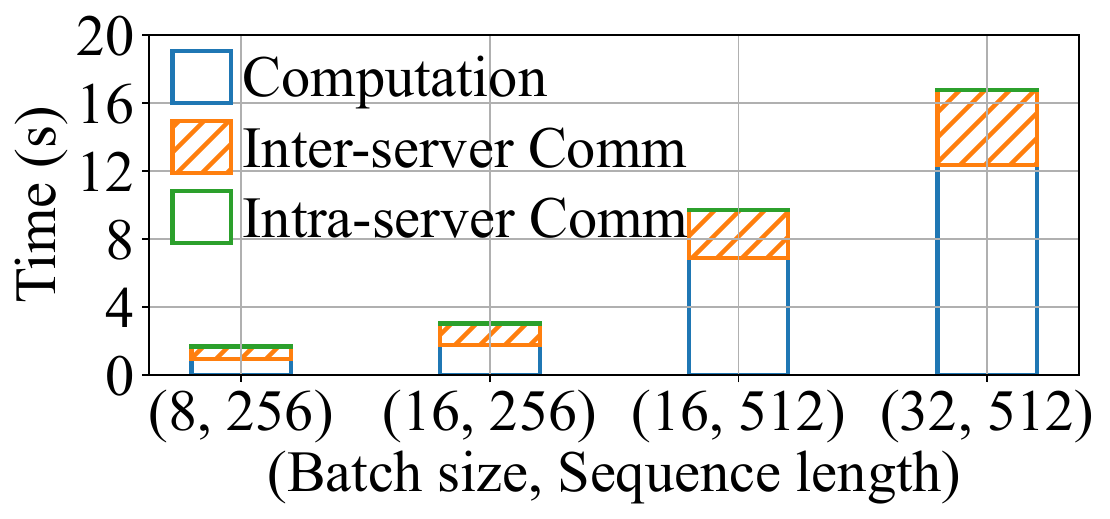}
    \vspace{-4mm}
    \caption{Components of inference latency of generating one token using LLaMA 7B in \sys.}
    \vspace{-6mm}
    \label{fig:latency_components}
\end{figure}

\noindent\textbf{C3: Mitigate communication overhead.} 
Versatile resource allocation of finer-grained \blocks provides several benefits for online serving, as mentioned in~\Cref{subsec:block_serving}. Yet \sys incurs additional non-negligible communication overhead. Specifically, each time a request batch is directed to a different device, it necessitates intra-server or inter-server communication. This communication might act as a bottleneck for the batch, directly contributing to increased latency. The communication overhead also linearly grows with the number of \block instances passed by a request batch during its lifecycle. Furthermore, the volume of data transferred escalates with the use of larger batch sizes per \block. Figure~\ref{fig:latency_components} breaks down the latency components involved in generating one new token across various request batches. Without any optimization to the communication, for a request batch of size 32 and length 512, the computation constitutes 62.9\% of the total latency, whereas the inter-server communication accounts for 36.4\%. 
% \aditya{is the figure with or without our optimizations for communication? be clear?}

%% file: modelzoo.tex
\section{\Block Zoo}
\label{sec:blockzoo}
In this section, we mainly answer three questions:
\begin{itemize}[noitemsep,topsep=0pt,leftmargin=*]
    \item How to evaluate equivalence among model components? (\textbf{C1}, \Cref{subsec:equivalence})
    \item What are the accuracy implications of \block sharing? (\Cref{subsec:accuracy_impl})
    \item How to partition LLMs into \blocks? (\Cref{subsec:partition})
    \item How to handle request routing between two equivalent \blocks with different embedding sizes? (\textbf{C1}, \Cref{subsec:stitching})
\end{itemize}

\subsection{Equivalence}
\label{subsec:equivalence}

% Our goal is to find such equivalence so that some of the \FPFT model components can be shared as well once they are partitioned into \blocks. We consider two scenarios here.
\begin{figure}[t]
    \centering
    \begin{minipage}{.46\columnwidth}
        \includegraphics[width=\columnwidth]{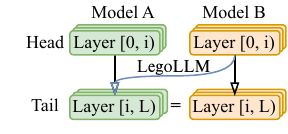}
       \vspace{-6mm}
        \caption{Routing requests when equivalence is found.}
        \label{fig:same_size_equivalence}
    \end{minipage}%
    \hfill
    \begin{minipage}{.46\columnwidth}
        \includegraphics[width=\columnwidth]{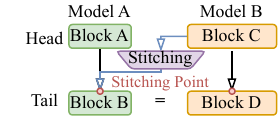}
        \vspace{-5mm}
        \caption{Stitch \blocks in different embedding sizes.}
        \label{fig:diff_size_equivalence}
    \end{minipage}
    \vspace{-3mm}
\end{figure}

% \noindent\textbf{Models with an identical architecture.} When two models share an identical architecture, we can establish the equivalence to optimize request routing. For instance, if tail layers (Layer $i$ to $L$) are determined to be equivalent, as depicted in Figure~\ref{fig:same_size_equivalence}, we have the flexibility to route requests from head layers of Model B to tail layers of Model A, which becomes particularly useful when the tail layers of Model B is experiencing high load. This strategy is also beneficial in conserving resources and reducing loading overhead when no instance of the tail layer of Model B is actively deployed.

% In this case, the degree of equivalence between model components can be systematically quantified. In \sys, we employ the canonical metric, cosine similarity, to evaluate the parametric similarity of each Transformer layer. 
% The similarity is defined as the weighted average of cosine similarities of all constituent parameters within each Transformer layer. That is, for layer $A_i$ and $B_i$ is:
% \begin{equation*}
%     \vspace{-1mm}
%     Eq(A_i, B_i) = \sum_{p = 1}^{n} (s(A_i^p) \times cos(A_i^p, B_i^p))/\sum_{p = 1}^{n}s(A_i^p),
%     \vspace{-1mm}
% \end{equation*}
% \bd{just move this to Section 2.2?}
% where $n$ is the number of parameters in $A_i$, $s(A_i^p)$ denotes the number of values in the $p$-th parameter in $A_i$, and $cos(A_i^p, B_i^p)$ is the cosine similarity between the $p$-th parameters of $A_i$ and $B_i$. A higher value indicates a high similarity or equivalence between two Transformer layers.

\noindent \PEFT models inherently possess equivalent components for reuse. We focus on verifying the equivalence among \FPFT model components. Gvien that different models may produce outputs of varying sizes (e.g. the embedding sizes of the 7B and 13B LLaMA models are 4096 and 5120, respectively), we utilize \textit{output vocabulary probabilities} to determine equivalence. By feeding the same data into the model, we transform the output of each Transformer layer into vocabulary probabilities and compute the cosine similarity of these probabilities as an indicator of equivalence. Figure~\ref{fig:output_cos_similarity} illustrates the output cosine similarity of
% Transformer
layers partitioned from models of different sizes. Since LLaMA 7B and LLaMA 13B have different number of transfomer layers, we group 3 layers from LLaMA 7B and 4 layers from LLaMA 13B into a single block for comparison. The average similarity is 0.9841 for LLaMA 7B and LLaMA 13B, and 0.998 for LLaMA 7B and Vicuna 7B.

%\ml{what is the TF layer group? Since the depths of layers (\# of transformers) are different across models (e.g., 7B vs 13B), I am guessing that a certain number of layers from a bigger model forms a group, whose output is compared with the output from a  single layer of a smaller model? If so, should we briefly define/explain what the group means or how layers are put into a group? or may using block rather than group be better?}

%\bd{use equivalence analysis to motivate our model switch strategy}

\subsection{Accuracy Implications}
\label{subsec:accuracy_impl}

\begin{figure}[t] 
    \begin{subfigure}{0.48\columnwidth}
        \includegraphics[width=\columnwidth]{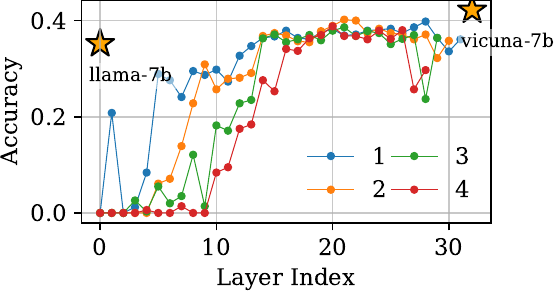}
        \centering
        \vspace{-4mm}
        \caption{LLaMA 7B and Vicuna 7B.}
        \label{fig:llama_vicuna_accs}
    \end{subfigure}
    \hfill
    \begin{subfigure}{0.48\columnwidth}
        \includegraphics[width=\columnwidth]{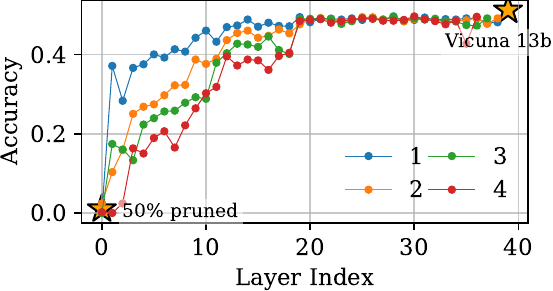}
        \centering
        \vspace{-4mm}
        \caption{Vicuna 13B w/ 50\% pruning.}
        \label{fig:vicuna_prune_50}
    \end{subfigure}
    \caption{Block Sharing Accuracy.}
    \label{fig:vicuna_mx}
    \vspace{-4mm}
\end{figure}

\begin{figure}[t]
    \begin{subfigure}{0.48\columnwidth}
        \includegraphics[width=\columnwidth]{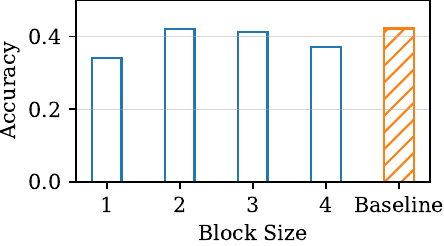}
        \vspace{-4mm}
        \centering
        \caption{LLaMA \& Vicuna 7B switching.}
        \label{fig:llama_vicuna_switch}
    \end{subfigure}
    \hfill
    \begin{subfigure}{0.48\columnwidth}
        \includegraphics[width=\columnwidth]{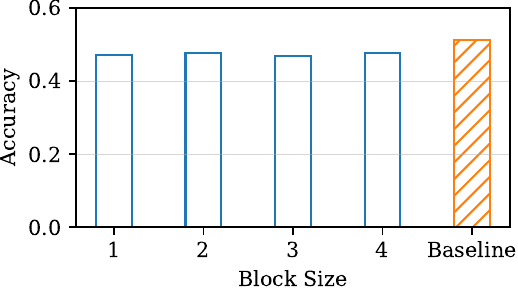}
        \vspace{-4mm}
        \centering
        \caption{Vicuna 13B block switching.}
        \label{fig:vicuna_switch}
    \end{subfigure}
    \caption{Block Switching Accuracy.}
    \label{fig:block_share_acc}
    \vspace{-7mm}
\end{figure}

\noindent\textbf{Block Sharing.} The output equivalence between foundational models and their \FPFT models, as well as between differently-sized foundational models, suggests potential for dividing models into sharable blocks. However, naively sharing these blocks may impact inference quality, even with high output similarity.

Figure~\ref{fig:llama_vicuna_accs} shows how block sharing affects accuracy. We treated each decoding layer as a single block and tested various numbers of shared blocks at different layer positions. For instance, with four blocks, we used layers 0-3 of LLaMA 7B and layers 4-31 of Vicuna 7B, then layers 1-4 of LLaMA 7B and layers 0, 5-31 of Vicuna 7B, and so on. The figure shows significant accuracy impact when sharing occurs at initial layers, but minimal impact at middle and end layers.

This trend is consistent even with different block configurations. Figure~\ref{fig:vicuna_prune_50} shows a similar pattern when applying 50\% parameter pruning to blocks. Unlike the previous experiment, we only use blocks from Vicuna 13B, and applied 50\% parameter pruning to selected blocks. The results indicate that pruning near the middle and end blocks has minimal accuracy impact, while initial blocks are less affected. Therefore, \block sharing should be prioritized for middle or end components before initial ones. This phenomenon has also been observed in previous studies~\cite{shortgpt,ffn-skip,skipdecode,bian-etal-2021-attention}, likely due to redundancy in later layers.

\noindent\textbf{Block Switching.} The design of \sys allows a request to be processed by an arbitrary block at any point, which may result in a single request frequently switching between blocks and potentially affecting inference quality. Figure~\ref{fig:llama_vicuna_switch} shows the impact of frequent block switching on accuracy. We selected layer 16 as the starting point for block switching due to stable accuracy and treated $x$ subsequent layers as a single block. For each block after layer 16, we continuously switch blocks between two models.
For instance,
% if the switching frequency is 2,
if $x = 2$, a request will use layers 17-18 from Vicuna 7B, then switch to layers 19-20 of LLaMA 7B, then switch back to layers 21-22 of Vicuna 7B, and so on. We applied a similar switching strategy to 50\% pruned Vicuna 7B blocks. Figures~\ref{fig:llama_vicuna_switch} and~\ref{fig:vicuna_switch} show that frequent block switching near the end has minimal impact on accuracy. In practice, the switching does not occur so frequently due to exceessive communication overhead.
% we do not expect the switching to occur so frequently due to exceessive communication overhead.

\subsection{Model Partitioning}
\label{subsec:partition}
With the evaluated equivalence and accuracy, \sys incorporates an automated mechanism that breaks down LLMs into \blocks. Each \block then becomes an independent engine during serving. 
% \hxx{this mechanism is not explained clearly enough in this section...}

\noindent\textbf{Principles of constructing \block zoo.} \sys follows two guiding principles: (1) avoid over-partitioning, (2) the preservation of architectural integrity. In this way, \sys can reuse the \blocks with minimal additional effort. %\ljm{minimum architecture -> attention, ffn, embedding, lm head}

The rationale for reducing redundancy is to prevent unnecessary duplication. If one model component has no variant, it is meaningless to over-partition it into excessively small \blocks. Such miniature partitioning is inefficient both in terms of computational power within each \block and due to the increased overhead from data transfer between \blocks.

The second principle directs \sys to partition the model only at clear architectural boundaries.
Based on \PEFT and \FPFT techniques we have adopted~\cite{alpaca2023,vicuna2023,peft,he2021towards}, \sys sets the following network components as the finest-grained network components that can compose a \block: $attention$, $ffn$, $embedding$, $lm\_head$. 
% \hxx{for systems folks, might be a good idea to explain this a bit more with a simple example. can we move the LoRA example up here to do this job? or you just mean to partition at layer boundaries?} 
This approach ensures that the output of one \block seamlessly serves as the input to another, facilitating a streamlined chain of \blocks for each request. \sys avoids intricate arithmetic operations between \blocks, simplifying the challenge of assembling them during the serving process. An illustrative example is the application of LoRA to two Linear operations within an Attention layer. Isolating LoRA as a separate \block would lead to the partitioning of the Attention layer into five \blocks, requiring \sys to sum up the outputs of the LoRA \block and the foundation Attention \block. Therefore, in \sys, each Attention with LoRA layer should be a separate \block.
% \hxx{good, so then what's the right way of partitioning it? each LoRA-ed attention should be a separate block even when they share the same foundation attention?}
\begin{figure}
    \centering
    \includegraphics[width=\columnwidth]{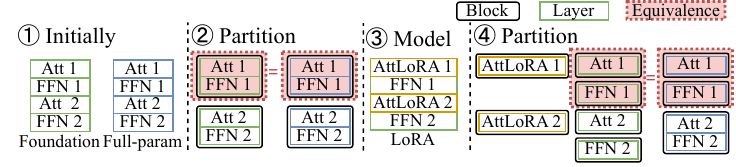}
    \vspace{-6mm}
    \caption{An example of \sys's lazy partitioning.}
    \label{fig:partition_example}
    \vspace{-6mm}
\end{figure}

\noindent\textbf{Lazy partitioning.} 
% \ljm{1. peft (remove current example), 2. full param -> same arch -> threshold -> refer to fig3, 3. full param -> diff arch -> perf tradeoff (i will add fig)} 
\sys employs a lazy partitioning strategy for all fine-tuned models utilizing a foundation model, slicing models as needed without preset restrictions on the size or architecture of the \blocks. For \PEFT models, it is partitioned into two \blocks, the foundation model and its \PEFT adapter. 

As for \FPFT models, after \sys evaluates equivalence between model components (\Cref{subsec:equivalence}), we perform lazy partitioning. As shown in Figure~\ref{fig:partition_example}, for an \FPFT model adapted from the foundation model (\circled{1}), \sys sets a threshold $T$ and group \textit{all connected network components} (e.g., transformer layer 1 from the fine-tuned model and the foundation) that have equivalence exceeding $T$ into a \block (\circled{2}). This action results in 4 \blocks, with two \blocks being considered equivalent. When a newly added \PEFT model (e.g., LoRA) arrives (\circled{3}), \sys (1) preserve its \PEFT adapters (e.g., AttLoRA1, AttLoRA2), (2) find all existing \blocks that share parameters (e.g., FFN1, FFN2) with the new model and (3) attempts to partition the existing \blocks to enable \block reuse. LoRA here has a different Attention layer. Therefore, we \textit{further} partition the foundation model into four \blocks so that they share the FFN layers in separately-provision \blocks (\circled{4}). This action results in 9 \blocks with two chain-of-blocks being equivalent. 

It is important to note that \sys uses \blocks as logical basic units of resource provisioning, and the granularity of \blocks does not necessarily introduce overhead during serving time: \sys considers overhead of pipelining \blocks at fine-granularity and could choose to allocate consecutive chain of \blocks in a single agent. The engine executes these \blocks as a single network sub-component. 

\subsection{Stitching \Block}
\label{subsec:stitching}

\begin{figure}[t]
    \centering
    \includegraphics[width=0.7\columnwidth]{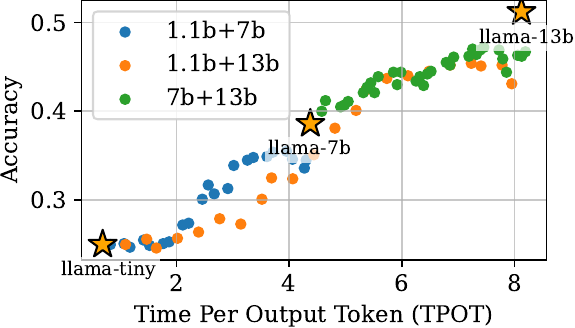}
    \vspace{-3mm}
    \caption{Accuracy \& TPOT after applying a stitching layer to LLaMA (1.1B, 7B, 13B)}
    \vspace{-3mm}
    \label{fig:stitch_llama}
\end{figure}

While requests can be directly routed between \blocks with the same embedding sizes, this does not apply to \blocks of different embedding sizes. When two \blocks of different embedding sizes are considered equivalent, such as \Blocks B and D illustrated in Figure~\ref{fig:diff_size_equivalence}, routing requests originally intended for \Block B to D presents a unique challenge. 

\noindent\textbf{Stitching \blocks.} We propose to use a stitching \block, a concept that has been recently explored~\cite{pan2023stitchable}. We follow the principle of fine-tuning LLMs to obtain a stitching \block. However, training a unique stitching \block for each potential \block pair is impractical. Thus, our objective is to design a \textit{generalizable} stitching \block. When multiple equivalences are identified between \blocks originating from the same two foundation models, requests can be seamlessly redirected at any stitching point using the stitching \block. 
% \hx{<- i didn't understand this sentence...}
\begin{table}[t]
    \centering
    \resizebox{0.8\columnwidth}{!}{\begin{tabular}{@{}lcc@{}}
    \toprule
    Stitching \Block & GPU hours & LM Head Cosine Similarity \\ \midrule
    (2048, 4096)     &  2.01       &  0.9634                    \\
    (4096, 5120)     &  4.33       &  0.9732                    \\
    (5120, 4096)     &  4.84       &  0.9683                   \\
    (4096, 8192)     &  6.32       &  0.9798                    \\
    (5120, 8192)     &  5.85       &  0.9612                   \\ \bottomrule
    \end{tabular}}
    \vspace{-3mm}
    \caption{Costs of training a stitching \block in different sizes on A100 GPUs and output similarity with the large model. }
    \vspace{-4mm}
    \label{table:stitching_training_perf}
\end{table}

We use a Linear layer to serve as the stitching \block. This stitching \block is trained while keeping the parameters of other \blocks static. This approach is inspired by~\cite{pan2023stitchable}. To generalize the stitching \block, we encode the positional information of the stitching point as an extra dimension. The position value is the sum of the positions of the head \block and the tail \block in their original foundation model. During training, the stitching block is initially placed at a shallow layer and progressively moved to deeper ones. This approach follows the idea of reusing layers in~\cite{lym2019mini, yang2017deep}. Table~\ref{table:stitching_training_perf} shows the training costs and the performance of the stitched model compared with the large model. Figure~\ref{fig:stitch_llama} shows the accuracy and latency when applying stitching block between different decoding layers in LLaMA 1.1B, 7B, and 13B.

The underlying hypothesis of our approach is rooted in the Transformer-based architecture of LLMs: the output embedding from any Transformer layer can be interpreted as natural language. This characteristic presents the opportunity to engineer a stitching \block with the unique capability of translating the output embedding from one dimension to another. This strategy allows the integration of varied \block sizes.

%% file: onlineserving.tex
\section{Online Serving}

% \begin{figure}[t]
%     \centering
%     \includegraphics[width=0.8\columnwidth]{figures/accs_lats_stitch.pdf}
%     \vspace{-3mm}
%     \caption{Accuracy after applying the stitching block across various model blocks.}
%     \vspace{-3mm}
%     \label{fig:stitch_accs}
% \end{figure}

\label{sec:serving}
We now present \sys's online serving system. We first explain how \sys deals with stateful KV cache (\textbf{C2}, ~\Cref{subsec:statful}). Then, we discuss \sys's two strategies to mitigate the impact of communication overhead (\textbf{C3}, \Cref{subsec:speculative}, \Cref{subsec:schedule}).

% \begin{algorithm}
% \caption{KV Cache Migration}
% \label{alg:random_mod}
% \begin{algorithmic}[1]
% \Function{selectSequences}{$S$, $M$}
% \State $\textit{sortedS} \gets$ \texttt{sort($S$, ref(), resumeTime())}
% \State $\textit{selectedS} \gets$ $\{\}$
% \While{$M > 0$}
%     \If{\texttt{resumeTime($s_i$)} $\geq$ \texttt{estimateMig}($s_i$)}
%         \State \textit{selectedS}.\texttt{append}($s_i$)
%         \State $M$ $\gets M - \texttt{sizeof($s_i$)}$
%     \EndIf
% \EndWhile
% \If{$selectedS$ = $\{\}$}
%     \State $selectedS \gets SortedS[0]$
% \EndIf
% \State \Return{$selectedS$}
% \EndFunction
% \Function{migrate}{$S$, $M$}
% \State $\textit{mirateS} \gets \texttt{selectSequences}(S, M)$
% \State $recompId \gets 0$
% \State $copyId \gets n$
% \While{$recompId \neq copyId$}
% \State \texttt{recompute($s_{recompId}$)}
% \State \texttt{copy($s_{copyId}$)}
% \State $recompId \gets recompId + 1$
% \State $copyId \gets copyId - 1$
% \EndWhile
% \State \Return{\texttt{S}}
% \EndFunction
% \end{algorithmic}
% \end{algorithm}

\subsection{KV Cache Coordination}
\label{subsec:statful}
\noindent\textbf{Memory bandwidth-bound KV cache.}
 Efficient stateful coordination of the KV cache is crucial for auto-regressive LLMs in \sys, as memory bandwidth constraints on the KV cache have been identified as a significant bottleneck in numerous studies. Existing systems process one batch of requests at a time, weighing the trade-off between recalculating the KV matrices and caching them in device memory. This trade-off reaches a point---determined by factors such as device type, model architecture, and request sequence length---where caching becomes more efficient than recalculation. However, as request sequences lengthen, memory bandwidth constraints become a performance-bounding factor when loading the KV cache~\cite{kwon2023efficient}.

\noindent\textbf{I/O and recalculation cost.} As aforementioned in~\Cref{subsec:challenges}, \sys's design complicates the problem. The assumption that requests are consistently processed by the same \block instances no longer holds, making I/O costs for transferring KV caches between instances unavoidable.

To migrate the KV cache from device $d_i$ to $d_j$, we optimize the process by overlapping KV cache recomputation with copying, thereby minimizing migration time. Given the fully known context, we employ chunked prefilling for efficient recomputation. For sequences to be migrated, denoted as $S = {s_i}_{1}^{n}$, we begin by recomputing the KV cache from the start of sequence $s_1$ while simultaneously copying the cache starting from the end of sequence $s_n$. The process concludes when recomputation encounters a KV cache page that has already been copied, indicating the completion of migration.

This approach is chosen for two key reasons. First, each token's KV cache depends on the KV caches of all preceding tokens. Recomputing the KV cache from the beginning of a sequence ensures the accuracy of the entire cache. In contrast, copying can take place independently, without relying on preceding caches. Second, ongoing requests on the target device may introduce latency during recomputation. By employing chunked prefill, we improve GPU utilization and mitigate the impact of KV cache recomputation on other tasks.

\noindent\textbf{Proactive KV Cache Migration.} As \sys{} may redirect requests to blocks lacking the necessary KV caches, this can introduce additional migration latency. While this overhead cannot be completely eliminated, it can be mitigated by proactively migrating KV caches in advance, thereby removing it from the critical path.

To ensure that migration does not introduce latency, it is essential to predict whether the KV cache will be used before the migration completes. The feasibility of predicting KV cache usage has been demonstrated in \cite{infercept}. We adopt the method proposed in \cite{infercept} to estimate the interception time: $T_{INT} = t_{now} - t_{call}$, where $t_{now}$ is the current time updated for each iteration, and $t_{call}$ is the time when the last interception was initiated.

\noindent\textbf{Memory Efficiency.} Modern LLM inference serving systems support paged attention~\cite{kwon2023efficient}, a technique that partitions the KV cache into smaller pages. This approach eliminates the need to store the entire KV cache in contiguous memory and allows for the sharing of KV cache pages across multiple requests, thereby enhancing memory efficiency. However, dynamically routing requests to blocks that lack the required KV cache can result in the creation of new KV cache pages. Since each device maintains its own dedicated KV cache page table, generating the same KV cache page on a different device leads to the duplication of KV pages. This duplication, which otherwise would only require a single KV page with an incremented reference counter, undermines the advantages of memory sharing.

To prevent memory waste, we prioritize migrating pages referenced by fewer requests before those referenced by more. We denote all KV pages as $C = \{c_1, c_2, ... c_n\}$, where each $c_i$ represents the underlying consecutive KV pages of request $s_i$, and $n$ is the total number of requests tracked by the system. We use $ref(c_i)$ to calculate the total number of pages referenced by more than one request in sequence $s_i$. For KV cache pages $c_i \in C$, we have $ref(c_i) \leq ref(c_{i+1})$. If $ref(c_i) = ref(c_{i+1})$, then $resumeTime(c_i) <= resumeTime(c_{i+1})$, where $resumeTime(c_i)$ is the estimated time the pages $c_i$ will be reused by the intercepted request $s_i$. 
%\ml{is this part related to Algo1? There is no reference to the algo 1. We have to refer to algo1 and put some description about the algo.} \bd{should we just remove the algo altogether? there's enough redundancy between the text and the algo} \ml{if you want, we can add the algo in an appendix. looks like nsdi allows appendices.}

\noindent\textbf{Prioritize KV cache owner.} Transferring KV caches to a new \block instance can introduce latency, impacting request processing times. This latency is subject to network bandwidth and varying network conditions. Although recalculating the KV cache can sometimes be more efficient than direct copying, it may still disrupt ongoing requests on the target device. Therefore, we prioritize \block instances that already possess the required KV cache before redirecting requests to a new device. This approach follows the principle of best-effort coordination. When candidate \block instances have the same status (e.g., queuing time), \sys’s agent prioritizes dispatching the request to the \block instance that holds the associated KV cache. In \Cref{subsec:schedule}, we provide a detailed discussion on how \sys'agent selects the appropriate \block instance.

% Comparing these two scenarios, it becomes evident that transferring the KV cache to a new \block instance is consistently less efficient than returning to the original KV cache owner. As for recalculating KV on the new \block instance, it is also less efficient because caching typically outperforms recalculations, and network bandwidth is generally not as high-capacity as memory bandwidth. We follow this insight and perform best-effort coordination. As long as the candidate \block instances have the same status (i.e., queuing time), \sys's agent should give priority to dispatch the request to the \block instance that holds the associated KV cache. In~\Cref{subsec:schedule}, we discuss in detail how \sys's agent picks the suitable \block instance.

% \noindent\textbf{Ownership of KV cache.} \sys is designed to store a KV cache for multiple requests, making it essential to distinguish the ownership of each KV cache matrix. Each agent maintains a dictionary mapping requests to the addresses of their corresponding KV caches. When processing a batch of requests, an agent of \sys looks up this dictionary using the hashed request ID to retrieve the memory address of the relevant KV matrices. Additionally, multiple devices can hold a KV cache for a single request. To avoid this redundancy, the most recent copy is retained and \sys's scheduler periodically removes the others. When \sys's scheduler receives a completed request, it lets the agents possessing the KV cache of that request delete the associated memory.

\subsection{Speculative Execution}
\label{subsec:speculative}
\begin{figure}[t]
    \centering
    \includegraphics[width=0.8\columnwidth]{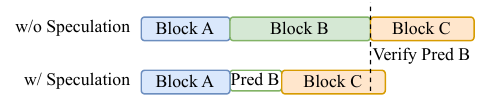}
    \vspace{-3mm}
    \caption{Speculative execution of \Block B.}
    \vspace{-3mm}
    \label{fig:spec_ideal}
\end{figure}

Request transfer overhead negatively impacts overall throughput in two significant ways. First, it directly and inevitably increases inference latency due to its inherent blocking nature: each subsequent block's operation depends on receiving output from the preceding block, creating a strong dependency. Second, it depletes network resources, as each redirection of a request to a different server involves network transfer. In \sys, we address the first issue with speculative execution and the second with \block scheduling and placement.

\noindent\textbf{Reduce latency via ahead-of-time prediction.} Considerable efforts have been dedicated to enhancing P2P network transfers~\cite{rajasekaran2022congestion,zhu2015congestion}, and such optimizations could be leveraged in \sys to diminish inference latency. However, our approach seeks to offset any latency increases by accelerating the inference pipeline itself. Drawing from the design in the operating system, we exploit speculative execution to expedite the inference process. Instead of awaiting actual outputs from a preceding \block, it predicts the outcome, allowing subsequent \blocks to proceed based on these predictions. Should the predictions align with the actual results, this speculation can lead to reduced latency. As illustrated in Figure~\ref{fig:spec_ideal}, when \Block B adopts speculative execution using Pred B, the inference latency is shorter if it is verified to be a correct prediction.

The challenge of a successful speculation lies in both the time and the accuracy of the prediction. If the prediction takes a longer time than the actual computation, the benefits of speculation are nullified. If the accuracy of predictions is low, reliance on the actual output becomes inevitable to ensure computational correctness, rendering the speculative efforts both redundant and a potential drain on resources. 
\begin{table}[t]
\centering
\resizebox{\columnwidth}{!}{\begin{tabular}{@{}llcc@{}}
\toprule
\Block & Pruned Percentage & Cosine Similarity & Computation Speedup \\ \midrule
5th Attention        &    49.71\%     &    0.7        &   37.43$\times$    \\
5th FFN            &    49.83\%       &     0.77       &   1.71$\times$      \\
5th Transformer      &    49.79\%        &     0.94       &    22.91$\times$     \\
5th--7th Transformers   &    49.79\%         &      0.84      &   18.55$\times$      \\ \bottomrule
\end{tabular}}
\vspace{-3mm}
\caption{Performance of \blocks' surrogates in terms of prediction similarity and computation time. \Blocks are pruned from LLaMA 7B based on the structured pruning technique described in~\cite{ma2023llmpruner}. }
\vspace{-3mm}
\label{table:block_surrogate_perf}
\end{table}

\noindent\textbf{\Block surrogates.} We leverage insights from existing work and empirical studies, coupled with our \block zoo, to construct high-fidelity surrogates for the \blocks. The surrogates are used for prediction. First, we consider pruned models. One typical pruning technique is sparse models. While sparsity typically results in a reduction of computation FLOPs, it does not always correspond to a commensurate decrease in computation time. Hence, our focus is on pruning techniques that meaningfully cut down computation time. We implement a strategy inspired by existing literature~\cite{ma2023llmpruner} to construct surrogates for our \blocks. In particular, we perform structured pruning by selectively removing sub-structure that generates less critical impact towards the model output and then add fine-tuned LoRA for performance recovery after the pruning process. Table~\ref{table:block_surrogate_perf} shows the output similarity and computation time of four \blocks and their surrogates. Second, we exploit equivalent \blocks with different computation costs. For instance, a \block partitioned from a 7B LLaMA model can serve as an effective surrogate for a \block from a 30B LLaMA model, which is adopted by~\cite{miao2023specinfer}.
\begin{figure}[t]
    \centering
    \includegraphics[width=\columnwidth]{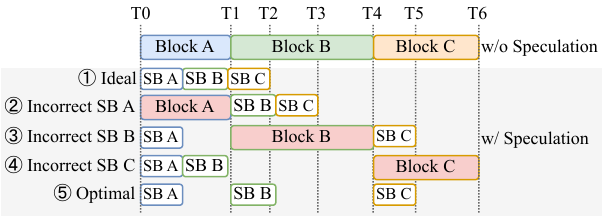}
    \vspace{-6mm}
    \caption{Possible scenarios when speculative execution is consecutively applied to all \blocks.}
    \vspace{-7mm}
    \label{fig:spec_cases}
\end{figure}

\noindent\textbf{Bottleneck-pinpointing speculation.} Naively applying speculative execution to blocks does not guarantee latency improvement, as it requires verifying the predicted results. In Figure~\ref{fig:spec_cases}, consider an inference pipeline composed of three \blocks where speculative execution is applied to each. Ideally, if all speculations are destined to be correct, the inference latency is expected to be reduced from T6 to T2 (\circled{1}). However, consecutive prediction verification can diminish these gains, especially when erroneous predictions happen at the latter part of the chain. For instance, if a prediction error arises at \Block C (last \block in the chain) (\circled{4}), the latency reduction would only be 0, compared to a reduction of T6-T3 if the error occurred earlier at \Block A (\circled{2}). Therefore, the optimal inference (\circled{5}) one could achieve is T5 instead of T2, because each speculative block needs to wait for the completion of the bigger verification block. In the worst case, where all three speculations are incorrect, there would be no latency reduction at all. Practically, due to the concurrent execution of \blocks and their surrogates, this could even result in additional slowdowns.

To avoid such outcomes, we follow two rules when implementing speculation. First, we avoid enabling speculation on consecutive \blocks to minimize accumulated errors and wasted resources. Second, speculation is not applied at the last \block in the chain because the final output, once relayed back to \sys's scheduler, is not correctable. Thus, during serving, we selectively apply speculation to the top-$k$ bottleneck \blocks---those characterized by their computation intensity or frequent use. The performance of speculative inference is evaluated in \Cref{subsec:components_perf}.

\subsection{Request and \Block Scheduling}
\label{subsec:schedule}

\noindent\textbf{Adaptive serving.} \sys's scheduler and agents perform adaptive serving when dispatching the request to a suitable \block instance. With the help of \block zoo, the candidate \block instances for each request include the \block that is explicitly listed in the chain and its equivalent \blocks. 

\sys's request scheduling strategy prioritizes \block instances that hold a request's KV cache, provided the device memory can accommodate the request data. If the device memory cannot hold the data, the strategy estimates the latency of each candidate instance and schedules the request to the instance with the lowest latency increase. When a \block does not have a request’s KV cache, the latency estimation for a candidate \block instance on device $c$ is as follows:

\begin{equation*}
    \begin{split}
    Latency_{d_c} &= T_{queue} + T_{compute} + T_{transfer} + T_{load}, \\
    T_{queue} &=  \sum_{i=1}^{n}{Comp(req_i)}, \\
    T_{compute} &= Comp(req), \\
    T_{transfer} &= \begin{cases} 
       T_{transfer}^{w/kv} & \text{agents dispatch,} \\
       \dfrac{D_{req}}{B_{net}(s, d_c)} & \text{scheduler dispatches,}
    \end{cases} \\
    %T_{load} &= \begin{cases} 
    %   max(\dfrac{D_{b^\prime}}{B_{mem}(d_c)} + \dfrac{D_{b}}{B_{net}(d_c)} - T_{transfer},\ 0)\ & \text{idle device,} \\
    %   \dfrac{D_{b^\prime}}{B_{mem}(d_c)} + \dfrac{D_{b}}{B_{net}(d_c)} & \text{loading overhead.}\\
    %\end{cases}
    T_{load} &= \begin{cases} 
       max(T_{swap} - T_{transfer},\ 0)\ & \text{idle device,} \\
       T_{swap} & \text{loading overhead,}\\
    \end{cases}\\
    T_{swap} &= \dfrac{D_{b^\prime}}{B_{mem}(d_c)} + \dfrac{D_{b}}{B_{net}(d_c)}.
    \end{split}
\end{equation*}
%\ml{revised equations; please revert if you don't like it}
We incorporate four key factors: queuing ($T_{queue}$), computation ($T_{compute}$), transfer ($T_{transfer}$), and the overhead associated with \block switching ($T_{load}$). %Queuing time 
$T_{queue}$ accounts for the duration required to process all $n$ queuing request batches.
%Transfer time
$T_{transfer}$ is subject to who initiates the dispatch: if done by the scheduler $s$, the request batch is at the start of the chain and no KV cache is involved. Otherwise, agent dispatching includes the costs of migrating the request and its KV cache (discussed in~\Cref{subsec:statful}). Here, $D_{req}$ is the size of the request token, and $B_{net}(s,\ d_j)$ denotes the network bandwidth between two devices. 
%The overhead from switching \block
$T_{load}$ depends on the current status of the device. If the device is busy, this loading overhead only includes $T_{swap}$, the time to move out the existing \block $b^\prime$ and load the new \block $b$. If idle, the \block's loading can be overlapped with other operations. Therefore, the loading overhead is $T_{swap} - T_{transfer}$. %includes the time to move out the existing \block $b^\prime$ and load the new \block $b$, minus the transfer overhead, 
% whichever is faster.
%\ml{made small edits; please revert if inappropriate/incorrect.}
%Here,
In $T_{swap}$, $B_{mem}(d_c)$ is the device memory bandwidth of candidate $d_c$; $D_{b^\prime}$ is the size of an existing block $b$. For simplicity, we keep \block inside device memory and leave \block swapping between other storage options for future work.

\noindent\textbf{\Block resource allocation.} \sys's scheduler determines the resource allocation of \blocks, similar to a traditional serving system. We discuss \sys's strategy for enabling independent per-block scaling and speculative execution. For scaling, we follow a straightforward policy centered on the queue length of the \block instance. If the queue length exceeds $t$\% of the maximum queue length (i.e. saturating all the device memory), we scale onto more devices starting from the most heavy-loaded \block instances. If the \block instance to be scaled possesses requests' KV cache, we balance the load by moving the state as well. As for speculative execution, \sys actively applies to the top-$k$ time-consuming \block instances to accelerate the inference pipeline. \sys deploys surrogates using a dedicated stream on the same device where the \blocks to be speculated are situated.

\noindent\textbf{Locality-aware \block placement.} \sys's second effort to mitigate the transfer overhead is the strategic placement of \blocks, aiming to reduce the reliance on network resources. During placement, \sys prioritizes locality---ensuring that \blocks with frequent direct inter-dependencies are close to each other. Ideally, these \blocks are placed on the same server. Such an arrangement leverages the full potential of high-capacity intra-server connections, such as NVLink interconnects, for the transfer of requests and KV cache, rather than resorting to the constrained inter-server links. 

In \sys, the locality is quantified by monitoring historical traffic, specifically by a counter recording the frequency of being directly inter-dependent between two \blocks. \Block pairs with a higher locality value are placed within the same server. Furthermore, \sys's scheduler dynamically adapts to evolving traffic patterns; should the observed locality change, \block instances are migrated as necessary to align with the new pattern. The benefit of locality-aware placement is discussed in \Cref{para:locality-aware-eval}.

%% file: implementation.tex
\section{Design Details}
\label{sec:impl}
We have implemented a prototype of \sys on top of vLLM. It is compatible with HuggingFace models. We use NCCL for data transfer among servers.

\noindent\textbf{Profiling.} To support the online serving system, \sys profiles \blocks by measuring computation time under different batch sizes, including surrogates and multiplexing performance. It also measures communication time between devices using NCCL primitives and the overhead of loading the \block engine from disk to host and device memory.
 
\noindent\textbf{Batching.} Computation efficiency improves with larger batch sizes, but enforcing a fixed large batch size complicates request reorganization.  Therefore, \sys loosely encourages batching within each \block instance. Upon receiving a new batch, \sys's agent queues it and attempts to pack it with neighboring requests, ensuring the combined batch does not exceed the upper batch size limit. If no requests are queued, the agent processes the batch directly. Requests reaching EOS are removed from the batch and sent to the scheduler.

\noindent\textbf{Request dispatching.} \sys's agents use a FIFO+priority queue, prioritizing requests that have left KV cache memory. Each \block instance maintains a countdown clock for auto-regressive requests, prioritizing their return. The scheduler and agents dispatch requests differently: agents determine candidate \blocks and pack them, broadcasting requests to available agents. The scheduler maintains a live record of \block placements, simplifying dispatching.

% \sys's agents adopt a FIFO+priority queue, where priority is assigned to requests that have left KV cache memory and may revisit in the future. Each \block instance maintains a countdown clock associated with every auto-regressive request based on the time estimation of one iteration, anticipating its potential return. As long as the countdown remains active, \sys's agent prioritizes the processing of returning requests. 

% The scheduler and agents dispatch the request in slightly different ways. \sys's agent needs to (1) determine the candidate \block for each request and pack them in different buckets and (2) obtain the placement of the \block candidates. \sys's agent initiates a broadcasting request, and those available agents hosting the candidates would respond. \sys's scheduler, on the other hand, keeps a live record of the placement of every \block instance in the cluster, thus easing the dispatching process. 

% \noindent\textbf{Offload to host memory.} Queuing requests and KV cache take up large spaces. In \sys, host memory is used if necessary to hold the data. In such a case, \sys's scheduler and agents consider the extra overhead of moving the data from host memory to device memory.

%% file: evaluation.tex
\section{Evaluation}
\label{sec:eval}
We evaluate \sys using testbed experiments. We seek to answer three main questions:
\begin{itemize}[noitemsep,topsep=0pt,leftmargin=*]
    \item What impact does \sys has on accuracy? (\Cref{subsec:overall_perf})
    \item Does \sys improve the overall throughput and resource utilization compared to existing solutions? (\Cref{subsec:overall_perf})
    \item For online serving, how much latency reduction is contributed by \sys's key designs? (\Cref{subsec:components_perf})
\end{itemize}
\subsection{Setup}
\label{subsec:setup}
\noindent\textbf{Cluster.} Our cluster has four servers, each of which is equipped with A100 GPUs with 80GB memory. Two servers have two GPUs, and the other two have four GPUs. The servers are interconnected with 100~Gbps links.

\noindent\textbf{LLM Models.} We use two foundation models in different sizes: 6B Chat-GLM~\cite{zeng2023glm-130b}, 7B LLaMA, and 13B LLaMA~\cite{touvron2023llama}. We consider fine-tuning techniques, including \FPFT (Vicuna~\cite{vicuna2023}) and \PEFT (Prefix-Tuning, Adapter, BitFit, and LoRA). We have 20 fined-tuned LLM models, each representing an application.

\begin{figure}[t]
    \centering
    \begin{minipage}[t]{.5\columnwidth}
      \centering
        \includegraphics[width=\columnwidth]{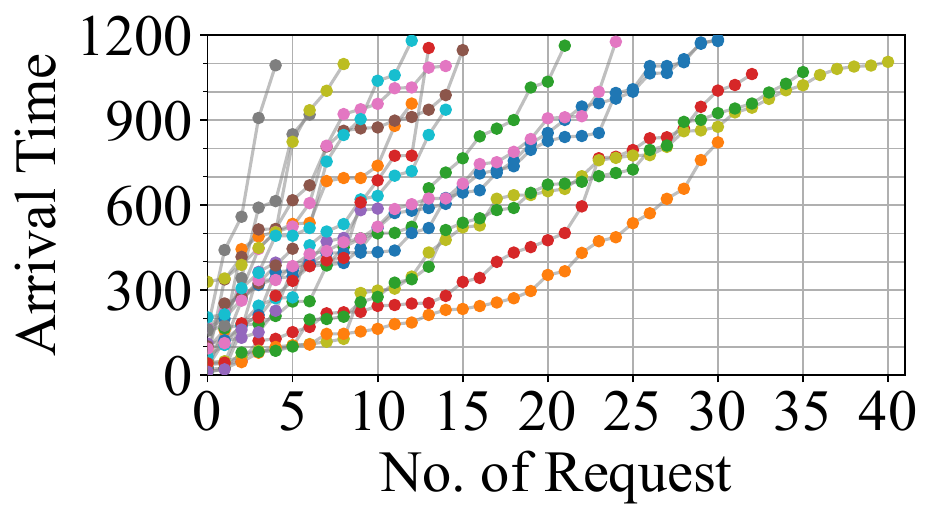}
        \vspace{-8mm}
        \caption{20-min workload.}
        \vspace{-4mm}
        \label{fig:synthetic_workload}
    \end{minipage}%
    \hfill
    \begin{minipage}[t]{.48\columnwidth}
         \includegraphics[width=\columnwidth]{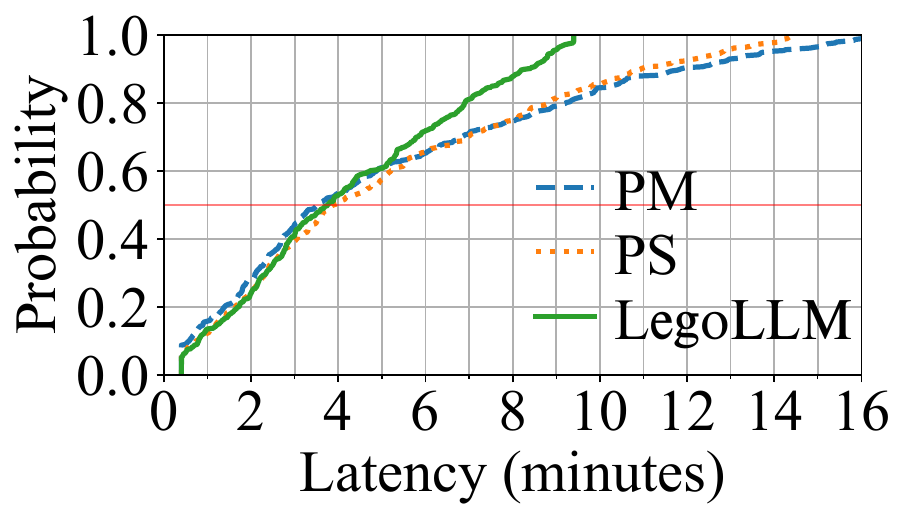}
       \vspace{-8mm}
        \caption{Latency CDF.}
        \label{fig:overall_latency_cdf}
    \end{minipage}
    \vspace{-4mm}
\end{figure}

\noindent\textbf{Workload.} We generate both production and synthetic workload traces (Figure~\ref{fig:synthetic_workload}) to evaluate \sys's performance. For synthetic traces, we use uniform distribution to determine the mean rate for each application. The mean rate serves as the weight to calculate the number of requests per application, creating varying popularity among applications. The arrival time of each request in one application is generated using a Poisson process with the given mean rate. The trace spans 20 minutes and includes 400 requests, following a generation approach similar to~\cite{sheng2023s}. For production traces, we use timing information from a Twitter trace collected over a month-long period~\cite{twitter_trace}. We set the minimum QPS to 1 and maximum QPS to 45. Previous work has shown this trace accurately reflects realistic inference workloads, exibiting diurnal patterns and unexpected spikes~\cite{mark_atc}.

\begin{figure}[t]
    \centering
    \includegraphics[width=0.7\columnwidth]{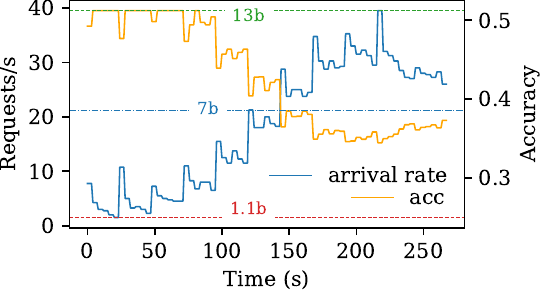}
    \vspace{-3mm}
    \caption{Accuracy and request arrival rate changes.}
    \vspace{-3mm}
    \label{fig:traces_accs}
\end{figure}

\noindent\textbf{Baselines.} We benchmark \sys with two baselines. %other solutions.
\begin{itemize}[noitemsep,topsep=0pt,leftmargin=*]
    \item Per-Model provisioning (PM): Each LLM is deployed independently. Models are scaled along with traffic demand.
    \item Parameter Sharing (PS): We merge \PEFT LLMs from the same foundation and deploy them as a complete unit, similar to S-LoRA~\cite{sheng2023s}. Since we support various fine-tuned applications, we use branching to differentiate each application. 
\end{itemize}

\noindent\textbf{Metrics.} We evaluate \sys with several metrics, including latency of serving one request, throughput (number of tokens per second), communication costs (percentage of time spent on data transferring), GPU utilization (SM efficiency), device memory consumption, and end-to-end accuracy. 

\noindent\textbf{\sys's configuration.} In this evaluation, we apply speculative execution to the top 10\% bottlenecked \block instances sorted by the time of completing their request queues. \sys's scheduler checks redundant KV cache every one minute. We set 0.95 as the cosine similarity threshold for determining whether \sys's surrogate prediction is accurate and 0.98 as the threshold for determining equivalence. The maximum sequence length allowed is 1024. \sys trains the stitching \block with the MMLU~\cite{hendrycks2020measuring}.

\subsection{Overall Performance}
\label{subsec:overall_perf}
We first evaluate the overall performance of \sys. The default number of applications is 20.

\begin{figure}[t]
    \centering
    \begin{minipage}[t]{.48\columnwidth}
        \includegraphics[width=\columnwidth]{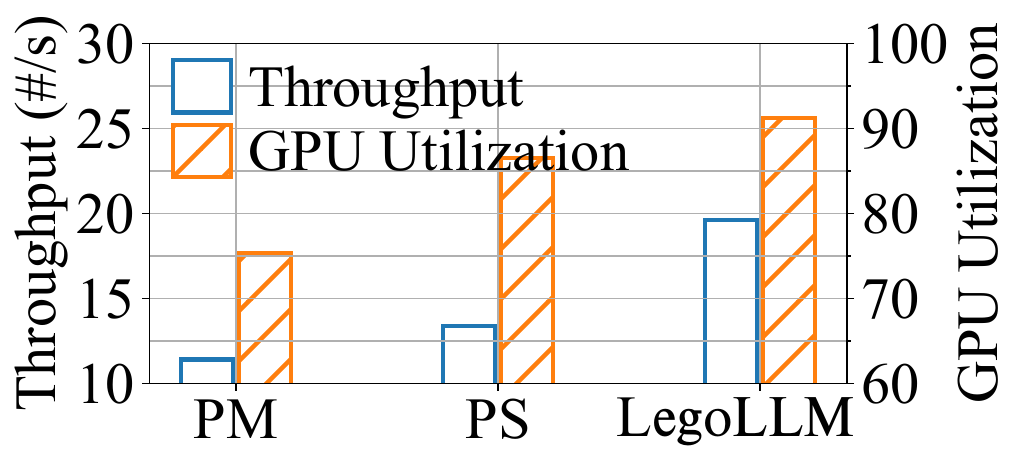}
        \vspace{-5mm}
        \caption{Throughput and GPU utilization.}
        \label{fig:overall_throughput_util}
    \end{minipage}%
    \hfill
    \begin{minipage}[t]{.48\columnwidth}
       \includegraphics[width=\columnwidth]{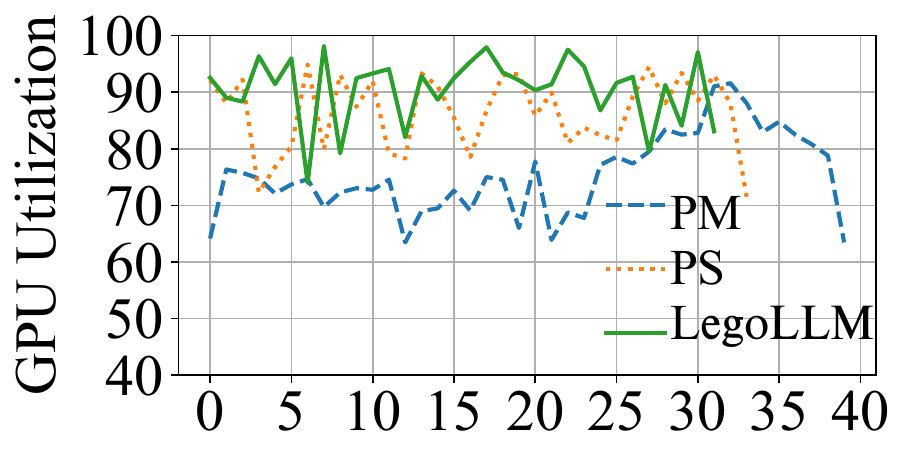}
       \vspace{-5mm}
        \caption{GPU utilization change over time.}
    \label{fig:util_time_series}
    \end{minipage}
    \vspace{-6mm}
\end{figure}

\noindent\textbf{Accuracy.} We use the production Twitter traces to study how accuracy reacts to changing request arrival rate. Figure~\ref{fig:traces_accs} illustrates the variation in accuracy over time under different 
%request
arrival rates. The models evaluated include LLaMA 1.1B, 7B, and 13B. Incoming requests prioritize the highest-accuracy blocks (13B) unless resources are unavailable. The average accuracy achieved is 42.1\%, which is 3.5\% higher than the closest option (7B). Higher request arrival rates cause \sys to redirect traffic to faster but less accurate blocks (1.1B) more frequently, resulting in a gradual decline in accuracy.

\noindent\textbf{Latency and throughput.} Figure~\ref{fig:overall_latency_cdf} depicts the CDF of the latency of completing a request in \sys. \sys's median latency is 3.34~min, comparable with PM (3.39~min) and PS (3.87~min). 
%provisioning.
The reduction of 95\%ile latency is more significant, 33.5\% and 23.4\% compared with PM and PS. The throughput of \sys is 1.71x of PM and 1.46x of PS
% provisioning
(Figure~\ref{fig:overall_throughput_util}). By decomposing models into more granular blocks, \sys enhances the efficiency of processing larger batch sizes. This approach significantly reduces tail latency, especially under high request rates. Though PS can serve various applications within one model, the cost of one inference process grows due to branching.
%\ml{any backup evidence on this claim?}
\begin{figure}[t]
    \centering
    \begin{minipage}[t]{.48\columnwidth}
        \includegraphics[width=\columnwidth]{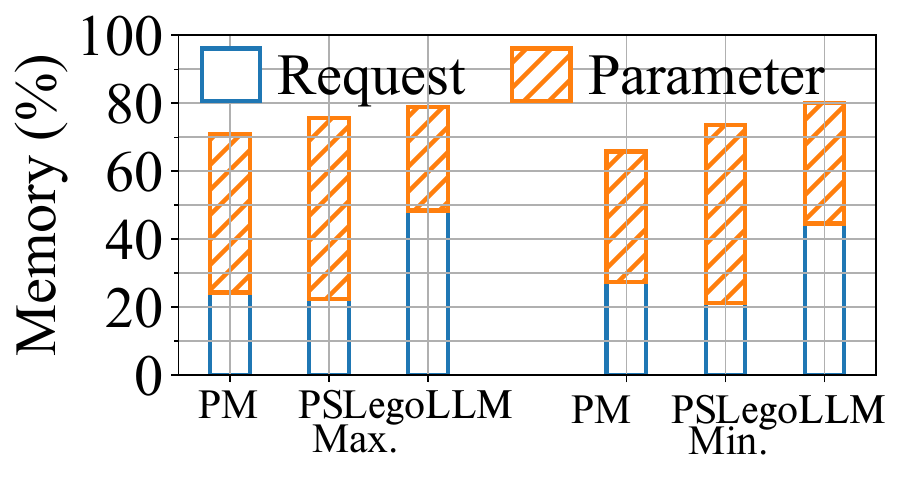}
        \vspace{-5mm}
        \caption{Memory usage of parameters and request-related data.}
        \label{fig:memory_usage_data_param}
    \end{minipage}%
    \hfill
    \begin{minipage}[t]{.48\columnwidth}
        \includegraphics[width=\columnwidth]{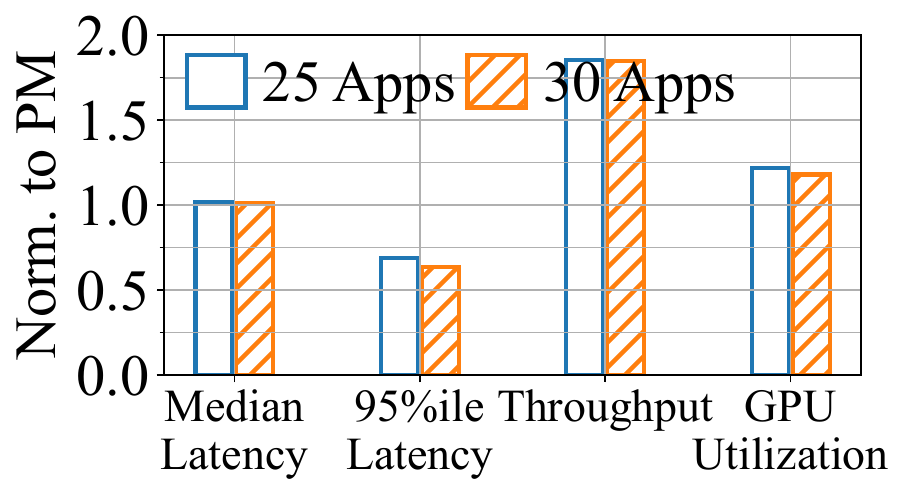}
        \vspace{-5mm}
        \caption{Performance changes when the number of applications grows.}
        \label{fig:overall_perf_no_app}
    \end{minipage}
    \vspace{-3mm}
\end{figure}

\noindent\textbf{GPU utilization.} We also measure overall GPU utilization indicated by SM efficiency and memory consumption in the cluster. We monitor the end-to-end serving process, and the average GPU utilization is improved by 20.1\% and 4.8\% compared with PM and PS. % provisioning.
Figure~\ref{fig:util_time_series} depicts the GPU utilization change over the serving process. With the help of adaptive serving, \sys efficiently dispatches requests under the existing deployment status to avoid frequent model loading and unloading. 
%As for memory consumption, 
We measure the memory consumption of model parameters and request-related matrices (including input, intermediate activations, output, and KV cache). In the best case, \sys takes 16.1\% less space on model parameters and 24.1\% more space on request-related matrices, indicating more data are being processed. 

\noindent\textbf{Number of applications.} We increase the number of applications from 10 to 30 with the same approach in~\Cref{subsec:setup}. When the number of applications grows to 30, and resources become increasingly constrained, \sys's performance gain is more prominent, achieving a 37.4\% reduction in 95\%ile latency and a 1.85x improvement in overall throughput (Figure~\ref{fig:overall_perf_no_app}). \sys exploits \block reuse to serve more applications under the same resource constraints and its flexibility brings efficiency in request serving.

\subsection{Effectiveness of \sys's Design}
\label{subsec:components_perf}
We then evaluate the effectiveness of \sys's key designs via an ablation study.
\begin{figure}[t]
    \centering
    \begin{minipage}[t]{.44\columnwidth}
       \includegraphics[width=\columnwidth]{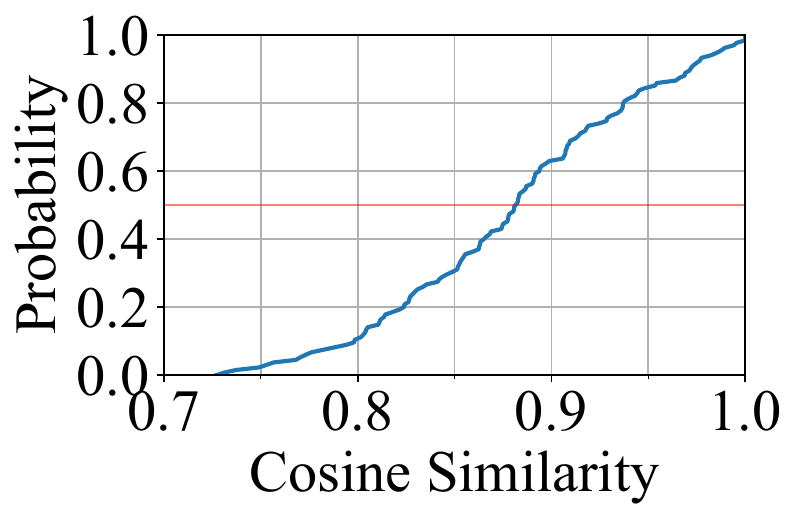}
       \vspace{-5mm}
        \caption{Cosine similarity of output vocabulary using adaptive serving and predetermined \block chains.}
    \label{fig:adaptive_serving_similarity}
    \end{minipage}%
    \hfill
    \begin{minipage}[t]{.52\columnwidth}
        \includegraphics[width=\columnwidth]{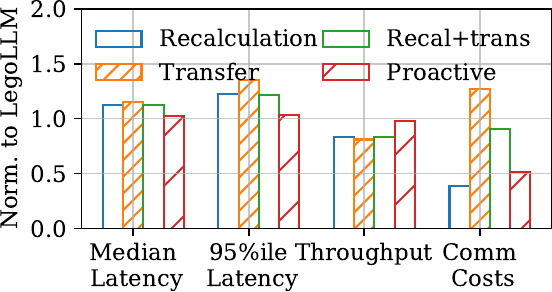}
        \vspace{-5mm}
        \caption{Ablation study of KV cache coordination strategy.}
        \label{fig:kv_cache_ablation}
    \end{minipage}
    \vspace{-5mm}
\end{figure}
\begin{figure}[t]
    \centering
    \begin{minipage}[t]{.42\columnwidth}
       \includegraphics[width=\columnwidth]{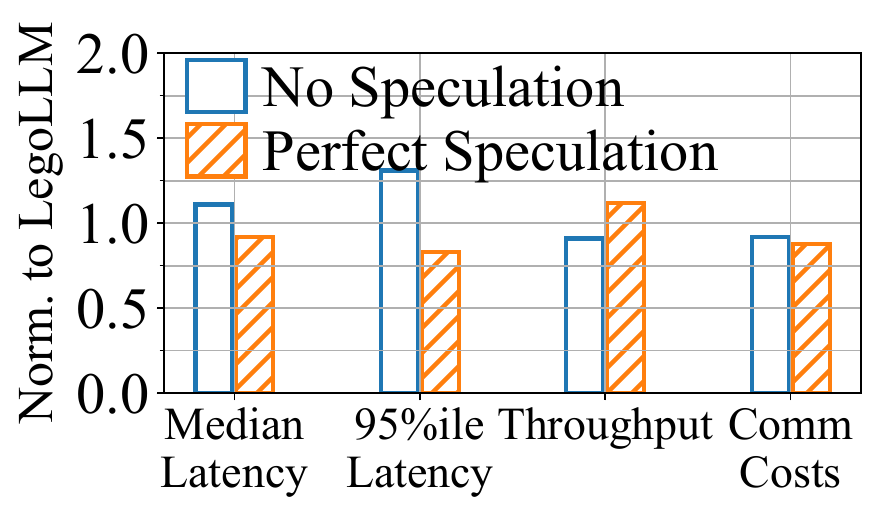}
       \vspace{-7mm}
        \caption{Ablation study of speculation: no speculation and perfect speculation.}
    \label{fig:speculation_ablation}
    \end{minipage}%
    \hfill
    \begin{minipage}[t]{.54\columnwidth}
        \includegraphics[width=\columnwidth]{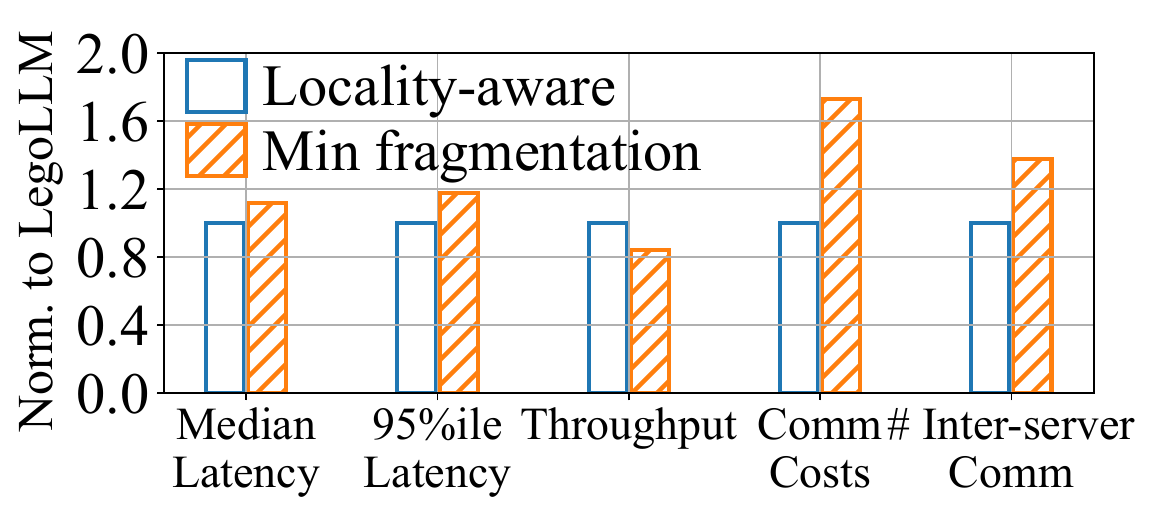}
        \vspace{-7mm}
        \caption{\sys compared with fragmentation-minimized placement.}
        \label{fig:placement_ablation}
    \end{minipage}
    \vspace{-6mm}
\end{figure}

\noindent\textbf{Adaptive serving.} We summarize the statistics of adaptive serving in \sys. When all the requests allow adaptive servings, 136 out of 400 requests are served with adaptive chains of \blocks. We compare the output vocabulary probabilities of these requests to the ones without adaptive serving and depict the cosine similarity CDF in Figure~\ref{fig:adaptive_serving_similarity}. The average similarity is 0.88.
%\ml{a minor point; from the figure, the median seems 0.88; is th median the same as the average?}
We disable adaptive serving in \sys and the 95\%ile latency and throughput are degraded by 15.6\% and 23.7\%, respectively. 
 
\noindent\textbf{Best-effort KV cache coordination.} 
%\ml{This para doesn't cover chunk-based prefill technique, etc. are we planning to include new results about the algorithm 1?} \bd{Updated Figure 24}
\sys performs best-effort dispatching by prioritizing the device with its KV cache memory. We consider two other solutions to verify if \sys's strategy is efficient: (1) All the KV cache are obtained using recalculation. (2) We always route the request back to the least busy device and let the KV cache owner transfer the cache to the instance. Figure~\ref{fig:kv_cache_ablation} shows the median and 95\%ile latency, throughput, and communication costs normalized to \sys. The 95\%ile latency is increased by 1.23x using recalculation and by 1.36x using least-busy-device routing. The communication costs are reduced significantly to 0.36 of \sys when using recalculation and increased to 1.28x when using least-busy-device routing.

\noindent\textbf{Speculative execution.} We evaluate the effectiveness of \sys's speculative execution via two approaches in Figure~\ref{fig:speculation_ablation}. First, we disable speculative execution and compare \sys's performance change. The median and 95\%ile latency inflates by 11.3\% and 31.6\%, compared with \sys with speculative execution. In this evaluation, speculative execution has been performed 231 times, 192 out of which have made predictions accurate enough to avoid correction. Second, we replace our surrogates with pseudo surrogates, where we assume their predictions are always accurate with an ideal 1/50 computation time of the speculated \block. The ideal 95\%ile latency and throughput are 87.3\% and 112.8\% of \sys with real surrogates. This shows that our surrogates are both efficient and accurate.

\noindent\textbf{Locality-aware \block placement.}\label{para:locality-aware-eval} We compare the communication costs between \sys's locality-aware placement and the widely adopted fragmentation-minimized (frag-min) placement. Figure~\ref{fig:placement_ablation} shows the average performance change of using 
%fragmentation-minimized
the frag-min placement. The median and 95\%ile latency is increased by 12.6\% and 18.2\%. The communication costs of one request sum up all the transfer costs, therefore presenting a significant inflation of 73.4\%. The locality-aware placement has reduced 72.3\% inter-server communications compared with the 
%fragmentation-minimized
frag-min placement strategy. 

%% file: discussion.tex
\section{Discussion}
\label{sec:discussion}
\noindent\textbf{Stitching models in different sizes.} The concept of model concatenation, particularly across varying sizes, has been investigated in prior research with notable efforts in Convolutional Neural Networks and Transformers~\cite{fang2023cooperative, pan2023stitchable}. In \sys, we introduce a simple technique to validate the potential of stitching \blocks of disparate sizes, ensuring that such integration is only executed between \blocks verified to be equivalent. However, it has limitations; it typically requires retraining and the maintenance of the original layer order. To overcome these limitations, a more advanced strategy is imperative based on a deeper comprehension of the models' underlying representations and functionalities.

\noindent\textbf{Other opportunities.} \sys's design brings other practical benefits. (1) Partial updates. Parameter updates can be applied in \block-granularity as well. With actor model~\cite{hewitt2010actor, mai2018chi}, updates can be performed concurrently with the serving process without interruption. Given the update frequency difference between the foundation and fine-tuning models, partial updates could effectively upgrade the system whenever necessary. (2) Early-exit~\cite{teerapittayanon2016branchynet} is a promising approach to reducing computation costs. However, batching limits its potential because the requests requiring the deepest computation always bottleneck the process. \sys's design choices naturally fit with early-exit, where requests can jump out of the batch at any \block, which is similar to DVABatch~\cite{cui2022dvabatch}, which tries to enable early-exit via multi-entry multi-exit batching.

%% file: related.tex
\section{Related Work}
\label{sec:related}
\begin{comment}
\noindent\textbf{General deep learning serving.} Extensive work has been focused on serving deep learning models. Clipper~\cite{crankshaw2017clipper}, TensorFlow Serving~\cite{tfserving}, and Clockwork~\cite{gujarati2020serving} explore the opportunity of batching requests and design efficient solutions for scheduling and placement. REEF~\cite{reef} and Shepherd~\cite{zhang2023shepherd} enable preemption for serving. AlpaServe~\cite{li2023alpaserve} shows the superior performances of applying model parallelism for multiplexing during serving. These efforts can be adapted to serve LLM workloads but do not provide tailored solutions. 
\end{comment}

\noindent\textbf{System for LLM models.} Model-internal optimization including FasterTransformer~\cite{FasterTransformer}, PagedAttention~\cite{kwon2023efficient}, FlashAttention~\cite{dao2023flashattention} and FlexGen~\cite{sheng2023flexgen} are complimentary to \sys's contributions. Orca~\cite{yu2022orca} introduces iteration-level scheduling to handle the sequence length difference among multiple requests. SpotServe~\cite{miao2023spotserve} enables LLM serving on preemptible instances with minimized communication costs and stateful recovery.
% Orca and SpotServe can be integrated with \sys's scheduler and agent.
PetS~\cite{280684} unifies four representative \PEFT techniques, allowing simultaneous execution of requests from different applications, similar to Parameter Sharing in~\Cref{sec:eval}. ~\cite{mosel,proteus,infaas} investigate accuracy-scaling to adapt to fluctuating workloads.

\noindent\textbf{Auto-regressive.} The inherent auto-regressive nature of LLMs limits their efficiency, making parallel decoding a key research focus. SpecInfer~\cite{miao2023specinfer} and \cite{leviathan2023fast} address the issue with speculative execution, using smaller models to predict multiple tokens in advance. \cite{monea2023pass, fu2023lookahead}  improve this by eliminating surrogate models, using the existing model with a few look-ahead tokens. These strategies generate multiple drafts to speed up token production. On the other hand, \sys applies speculative execution at the \block-level in a selective manner, optimizing the generation process at the individual token level. 

%% file: conclusion.tex
\section{Conclusion}
\label{sec:conclusion}
We present \sys, a multi-tenant finer-grained serving system tailored for LLM workloads. In \sys, we show the effectiveness of improving throughput by allowing model component reuse with \blocks. We enable adaptive serving, effectively coordinate multiple requests' KV cache, and mitigate the communication costs to improve serving efficiency. Experiments show the efficiency of \sys. We plan to extend \sys to support more models and evaluate it on larger-scale clusters.